\documentclass[12pt]{iopart}
\usepackage{iopams,mathptm,times,graphicx}
\usepackage{subfigure}
\usepackage{subeqnarray}
\usepackage{array}
\begin{document}

\title[]{An integrable semi-discretization of the coupled Yajima--Oikawa system}

\author{Junchao Chen$^{1,2}$, Yong Chen$^{1}$, Bao-Feng Feng$^2$, Ken-ichi Maruno$^{3}$ and Yasuhiro Ohta$^4$
}
\address{$^1$~Shanghai Key Laboratory of Trustworthy Computing, East China Normal University,
Shanghai, 200062, People's Republic of China}
\address{
$^2$~Department of Mathematics,
The University of Texas-Rio Grande Valley,
Edinburg, TX 78541, USA}
\address{$^3$~Department of Applied Mathematics, School of Fundamental Science and Engineering,
Waseda University, 3-4-1 Okubo, Shinjuku-ku, Tokyo 169-8555, Japan
}
\address{$^4$~Department of Mathematics, Kobe University, Rokko, Kobe 657-8501, Japan}
\ead{ychen@sei.ecnu.edu.cn, baofeng.feng@utrgv.edu, kmaruno@waseda.jp,ohta@math.kobe-u.ac.jp}

\begin{abstract}
In the present paper, an integrable semi-discrete analogue of the one-dimensional coupled Yajima--Oikawa system, which is comprised of multicomponent short-wave and one component long-wave, is proposed by using Hirota's bilinear method. Based on the reductions of the B\"{a}cklund transformations of the semi-discrete BKP hierarchy, both the bright and dark soliton (for the short-wave components) solutions in terms of pfaffians are constructed.
\end{abstract}

\date{\today}

\maketitle

\section{Introduction}
Both the nonlinear Schr\"{o}dinger (NLS) equation and the Yajima-Oikawa (YO) equation are import models arising in a variety of physical contexts \cite{Ablowitzbook}. It is known that they can be derived from the so-called $k$-constrained KP hierarchy \cite{ChengY}. Specifically, the NLS equation corresponds to the $k$-constrained KP hierarchy with $k=1$, while the YO equation is associated with the case with $k=2$ \cite{LorisRW}.

The integrable discretization of nonlinear Schr\"{o}dinger equation
\begin{equation}
\mathrm{i}q_{n,t}=\left( 1+\sigma |q_{n}|^{2}\right) \left(
q_{n+1}+q_{n-1}\right) \,
\end{equation}
was originally derived by Ablowitz and Ladik  \cite{ablowitz1975nonlinear,ablowitz1976nonlinear}, and was named Ablowatiz-Ladik (AL) lattice. Similar to the continuous case, the AL lattice equation admits bright soliton solution for the focusing case ($\sigma =1$) \cite{TsujimotoBookchapter,Narita1990} and dark soliton solution for the defocusing case ($\sigma =-1)$ \cite{OhtaMaruno}. The geometric construction of the AL lattice was given by Doliwa and Santini \cite{DoliwaAL}.

The semi-discrete coupled nonlinear Schr\"{o}dinger equation
\begin{eqnarray}
\mathrm{i}q_{n,t}^{(1)} &=&\left( 1+\sigma _{1}|q_{n}^{(1)}|^{2}+\sigma
_{2}|q_{n}^{(2)}|^{2}\right) \left( q_{n+1}^{(1)}+q_{n-1}^{(1)}\right) \,, \\
\mathrm{i}q_{n,t}^{(2)} &=&\left( 1+\sigma _{1}|q_{n}^{(1)}|^{2}+\sigma
_{2}|q_{n}^{(2)}|^{2}\right) \left( q_{n+1}^{(2)}+q_{n-1}^{(2)}\right)
\end{eqnarray}
where $\sigma _{i}=\pm 1 (i=1,2)$, is of importance both mathematically and physically. It was solve by the inverse
scattering method in \cite{TsuchidasdCNLS,TsuchidadCNLS}.
The general multi-soliton solution in terms of pfaffians was found recently in \cite{ohta2009discretization,feng2015nbright},
which is of bright type for the focusing-focusing case ($%
\sigma _{1}=\sigma _{2}=1$),
is of dark type for the defocusing-defocusing case ($%
\sigma _{1}=\sigma _{2}=-1$),
and could be of mixed type for the focusing-defocusing case ($\sigma_1=1$, $\sigma_2=-1$).

In compared with the successful construction of the integrable discrete analogue of the NLS equation and the coupled NLS equations, the integrable discrete analogues of the YO equation and its multi-component generalizations are missing. It should be pointed out that fully discrete NLS and YO equations were constructed most recently \cite{RWHattori}, however, their semi-discrete limits cannot converge to the model such as the Ablowatiz-Ladik lattice.  Therefore, it is of important to construct an integrable semi-discrete analogue of the one-dimensional (1D) coupled Yajima-Oikawa (YO) system:
\begin{eqnarray}
\label{sdyo-01}&& {\rm i}S^{(\mu)}_t-S^{(\mu)}_{xx}-LS^{(\mu)}=0,\ \ \mu =1,2,\cdots, M, \\
\label{sdyo-02}&& L_t=2\sum^M_{\mu=1}c^{(\mu)} (|S^{(\mu)}|^2)_x,
\end{eqnarray}
where $c^{(\mu)}$ are arbitrary real constants, $S^{(\mu)}$ and $L$ indicate the $\mu$th short-wave (SW) and long-wave (LW) components, respectively.

%Hereafter, we refer to the above multicomponent system as the ($M$+1)-component coupled YO system.
When $M=1$, the original YO system described such a physical process,
in which a resonant interaction takes
place between a weakly dispersive LW and a SW
packet when the phase velocity of the former exactly
or almost matches the group velocity of the latter \cite{zakharov1972collapse}.
Yajima and Oikawa \cite{yajima1976formation} proposed the model equation for the
interaction of a Langmuir wave with an ion-acoustic wave in
a plasma and showed that the system is integrable by the inverse scattering transform method.
This model equation was also derived in diverse physical backgrounds such as hydrodynamics \cite{benny1977general}, nonlinear optics \cite{kivshar1992stable,chowdhury2008long} and biophysics \cite{davydov1991solitons}.
It admits both bright and dark soliton solutions \cite{ma1979some,ma1978complete,willox1999pseudo}.
The rogue wave solutions to the 1D YO system have
recently been derived by using Hirota's bilinear method
\cite{wing2013rogue} and Darboux transformation \cite{chen2014dark}.
For the multicomponent 1D YO system (\ref{sdyo-01})-(\ref{sdyo-02}), it can be obtained from the KP-hierarchy by reduction as shown in Ref. \cite{zhang1994solutions,sidorenko1993multicomponent,liu1996bi}.
Most recently, Kanna et al. \cite{kanna2013general} considered the multicomponent coupled 1D YO system (\ref{sdyo-01})-(\ref{sdyo-02}),
and pointed out that the coupled YO system with two component LWs ($M=2$)
can be deduced from a set of three coupled NLS equations governing the propagation of three optical fields in a triple mode optical fiber by applying the asymptotic reduction procedure \cite{kanna2013general}.
Eqs.(\ref{sdyo-01})-(\ref{sdyo-02}) has also been derived to describe the interaction between a quasi-resonance circularly polarized optical pulse and a long-wave electromagnetic spike \cite{sazonov2011vector}.
In the context of many-component magnon-phonon system, such a multicomponent YO system has also been proposed and
its corresponding Hamiltonian formalism was studied \cite{myrzakulov1986particle}.
Also, the authors in ref.\cite{kanna2013general} have carried out Painlev\'{e} analysis for Eqs.(\ref{sdyo-01})-(\ref{sdyo-02}) and  obtained the general bright N-soliton solution in the Gram determinant form.
Soon after, they constructed an extensive set of exact periodic solutions in terms of Lam\'{e} polynomials of order one and two \cite{khare2014elliptic}.
%The rogue waves of the coupled YO system with $M=2$ have been reported in Ref.\cite{chen2014coexisting}.
With the help of the KP-hierarchy reduction method, we have recently considered the dark soliton, mixed soliton and rational solutions for the coupled YO system (\ref{sdyo-01})-(\ref{sdyo-02}) in \cite{chen2015multi,chen2015general,chen2015rational}.

Integrable discretizations of integrable systems have received considerable attention recently (see, e.g., \cite{levi2000side} and references therein).
So far, several approaches have been developed to construct integrable difference analogues of soliton equations.
Ablowitz and Ladik proposed a method of integrable discretization by using Lax pairs \cite{ablowitz1975nonlinear,ablowitz1976nonlinear}.
Based on the bilinear form, Hirota developed a method to construct integrable discrete analogues of soliton equations \cite{hirota1977nonlinear1,hirota1977nonlinear2,hirota1977nonlinear3}.
Date et al \cite{date1982method1,date1982method2,date1983method3,date1983method4,date1983method5} have shown an effective way to discretize integrable equations via the transformation group theory, in which
many of discretization versions of soliton equations can be derived by reduction.
Suris \cite{suris2003problem} also developed a general
Hamiltonian approach for integrable discretizations of soliton equations.
%Among above mentioned methods, the bilinear integrable discretization method
%is a direct and elegant method.
In a series of recent research,
the bilinear integrable discretization method has been widely applied to construct discrete schemes of the coupled Nonlinear Schr{\"o}dinger equation \cite{ohta2009discretization,feng2015nbright}, the (2+1)-dimensional sinh-Gordon
equation \cite{hu2007integrable}, the two--dimensional Leznov lattice equation \cite{hu2005integrable}, the Camassa--Holm equation \cite{ohta2008integrable,feng2010self}, the short pulse equation \cite{feng2010integrable-sp} and coupled short pulse equation \cite{feng2015integrable-msp,feng2015integrable-csp}, the short-wave model of the Camassa--Holm equation \cite{feng2010integrable-sw}, the reduced Ostrovsky equation \cite{feng2015integrable}, the coupled integrable dispersionless equation \cite{vinet2013discrete},  the integrable (2+1)-dimensional Zakharov equation \cite{yu2015dynamics} and so on.

The goal of the present paper is to construct an integrable semi-discrete analogue of
the coupled YO system by virtue of Hirota's bilinear method, and derive both the bright and dark soliton (for the SW components) solutions by the KP-hierarchy reduction method.
The remainder of the paper is organized as follows.
In section 2, we present an integrable semi-discrete version of the coupled YO system.
In section 3 and 4, the bright and dark soliton solutions in terms of pfaffians of the semi-discrete coupled YO system are constructed based on two types of B\"{a}cklund transformation of the BKP hierarchy.
%We the dark soliton solution of the semi-discrete coupled YO system in section 4.
Section 5 is concluded by some comments and discussions.

\section{Integrable semi-discrete coupled YO system}
Through the dependent variable transformation
\begin{eqnarray}
\label{sdyo-03}S^{(\mu)}=\frac{G^{(\mu)}}{F},\ \ L=2(\ln F)_{xx},\ \ \mu =1,2,\cdots, M,
\end{eqnarray}
the 1D coupled YO equations (\ref{sdyo-01})-(\ref{sdyo-02}) can be cast into the bilinear form
\begin{eqnarray}
\label{sdyo-04}&& ({\rm i}D_t-D^2_x )G^{(\mu)} \cdot F=0,\ \ \mu =1,2,\cdots, M,\\
\label{sdyo-05}&& D_xD_t F \cdot F-2c F \cdot F=2\sum^M_{\mu=1} c^{(\mu)} G^{(\mu)}\bar{G}^{(\mu)},
\end{eqnarray}
where $c$ is an integral constant and $\bar{}$  means complex conjugate.
The Hirota's bilinear differential operator is defined by
\begin{eqnarray*}
D^n_xD^m_t(a\cdot b)=\bigg( \frac{\partial}{\partial x} - \frac{\partial}{\partial x'} \bigg)^n
\bigg( \frac{\partial}{\partial t} - \frac{\partial}{\partial t'} \bigg)^m
a(x,t)b(x',t')\bigg|_{x=x',t=t'} .
\end{eqnarray*}

By discretizing the spacial part of the above bilinear equations,
\begin{eqnarray}
\label{sdyo-06}&& D^2_xG^{(\mu)} \cdot F \rightarrow \frac{1}{\epsilon^2}(G^{(\mu)}_{n+1}F_{n-1}-2G^{(\mu)}_nF_n+G^{(\mu)}_{n-1}F_{n+1}),\\
\label{sdyo-07}&& F_x=\frac{1}{\epsilon}(F_{n+1}-F_{n}),
\end{eqnarray}
one can obtain
\begin{eqnarray}
\label{sdyo-08}&& {\rm i}D_tG^{(\mu)}_n \cdot F_n -\frac{1}{\epsilon^2}(G^{(\mu)}_{n+1}F_{n-1}-2G^{(\mu)}_nF_n+G^{(\mu)}_{n-1}F_{n+1})=0,\ \ \mu =1,2,\cdots, M,\\
\label{sdyo-09}&& \frac{1}{\epsilon }D_t F_{n+1} \cdot F_n-c F_n \cdot F_n=\sum^M_{\mu=1} c^{(\mu)} G^{(\mu)}_n\bar{G}^{(\mu)}_n.
\end{eqnarray}

Furthermore, we require that the discretized bilinear forms are invariant under the gauge transformation
\begin{eqnarray*}
&& F_n\rightarrow F_n\exp(q_0n),\ \  G^{(\mu)}_n\rightarrow G^{(\mu)}_n\exp(q_0n),
\end{eqnarray*}
then one gets the gauge invariant semi-discrete bilinear YO equations
\begin{eqnarray}
\label{sdyo-10}&&\hspace{-0.5cm} {\rm i}D_tG^{(\mu)}_n \cdot F_n -\frac{1}{\epsilon^2}(G^{(\mu)}_{n+1}F_{n-1}-2G^{(\mu)}_nF_n+G^{(\mu)}_{n-1}F_{n+1})=0,\ \ \mu =1,2,\cdots, M,\\
\label{sdyo-11}&&\hspace{-0.5cm} \frac{1}{\epsilon }D_t F_{n+1} \cdot F_n- c F_{n+1} \cdot F_n=\sum^M_{\mu=1} \frac{c^{(\mu)}}{2} (G^{(\mu)}_{n+1}\bar{G}^{(\mu)}_n+G^{(\mu)}_{n}\bar{G}^{(\mu)}_{n+1}).
\end{eqnarray}

Let
\begin{eqnarray}
\label{sdyo-12}&& S^{(\mu)}_n=\frac{G^{(\mu)}_n}{F_n},\ \  L_n=\frac{2}{\epsilon^2}\left( \frac{F_{n+1}F_{n-1}}{F^2_n}-1 \right),\ \
 \hat{\hat{L}}_{n}=2\ln F_n,
\end{eqnarray}
the bilinear equations (\ref{sdyo-10})-(\ref{sdyo-11}) are transformed into
\begin{eqnarray}
\label{sdyo-13}&&\hspace{-0.7cm} {\rm i}\frac{d}{dt}S^{(\mu)}_n -\frac{1}{\epsilon^2}(S^{(\mu)}_{n+1}+S^{(\mu)}_{n-1}-2S^{(\mu)}_n) -\frac{1}{2}L_n(S^{(\mu)}_{n+1}+S^{(\mu)}_{n-1})=0, \ \ \mu =1,2,\cdots, M, \\
\label{sdyo-14}&&\hspace{-0.7cm} \frac{1}{\epsilon}(\hat{\hat{L}}_{n+1,t}-\hat{\hat{L}}_{n,t})-2c
=\sum^M_{\mu=1} c^{(\mu)} (S^{(\mu)}_{n+1}\bar{S}^{(\mu)}_n+S^{(\mu)}_{n}\bar{S}^{(\mu)}_{n+1}).
\end{eqnarray}
By using the relation $\frac{1}{2}(\hat{\hat{L}}_{n+1}-2\hat{\hat{L}}_n+\hat{\hat{L}}_{n-1})=\ln(1+\frac{\epsilon^2}{2}L_n)$, we propose the following discrete system
\begin{eqnarray}
\label{sdyo-15}&&\hspace{-0.7cm} {\rm i}\frac{d}{dt}S^{(\mu)}_n -\frac{1}{\epsilon^2}(S^{(\mu)}_{n+1}+S^{(\mu)}_{n-1}-2S^{(\mu)}_n) -\frac{1}{2}L_n(S^{(\mu)}_{n+1}+S^{(\mu)}_{n-1})=0, \ \ \mu =1,2,\cdots, M, \\
\label{sdyo-16}&&\hspace{-0.7cm} 2\frac{d}{dt}\frac{\ln(1+\frac{\epsilon^2}{2}L_n)}{\epsilon^2}=
\sum^M_{\mu=1} \frac{c^{(\mu)}}{\epsilon} [(S^{(\mu)}_{n+1}-S^{(\mu)}_{n-1})\bar{S}^{(\mu)}_n+S^{(\mu)}_{n}(\bar{S}^{(\mu)}_{n+1}-\bar{S}^{(\mu)}_{n-1})],
\end{eqnarray}
which converges to the coupled YO system (\ref{sdyo-01})-(\ref{sdyo-02}) when $\epsilon\rightarrow 0$.

For simplicity, by taking $\epsilon=1$ and applying the gauge transformation $S^{(\mu)}_n\rightarrow \exp(2{\rm i}t)S^{(\mu)}_n$,
one can obtain the semi-discrete YO equations
\begin{eqnarray}
\label{sdyo-17}&& {\rm i}\frac{d}{dt}S^{(\mu)}_n =(S^{(\mu)}_{n+1}+S^{(\mu)}_{n-1})(1 +\frac{1}{2}L_n),\ \ \mu =1,2,\cdots, M,\\
\label{sdyo-18}&& \hat{\hat{L}}_{n+1,t}-\hat{\hat{L}}_{n,t}-2c
=\sum^M_{\mu=1} c^{(\mu)} (S^{(\mu)}_{n+1}\bar{S}^{(\mu)}_n+S^{(\mu)}_{n}\bar{S}^{(\mu)}_{n+1}),\\
\label{sdyo-19}&& \ln(1+\frac{ L_n}{2})=\frac{1}{2}(\hat{\hat{L}}_{n+1}-2\hat{\hat{L}}_n+\hat{\hat{L}}_{n-1}),
\end{eqnarray}
or
\begin{eqnarray}
\label{sdyo-20}&&{\rm i}\frac{d}{dt}S^{(\mu)}_n =(S^{(\mu)}_{n+1}+S^{(\mu)}_{n-1})(1 +\frac{1}{2}L_n),\ \ \mu =1,2,\cdots, M,\\
\label{sdyo-21}&& 2\frac{d}{dt}\ln(1+\frac{ L_n}{2})=
\sum^M_{\mu=1} c^{(\mu)} [(S^{(\mu)}_{n+1}-S^{(\mu)}_{n-1})\bar{S}^{(\mu)}_n+S^{(\mu)}_{n}(\bar{S}^{(\mu)}_{n+1}-\bar{S}^{(\mu)}_{n-1})].
\end{eqnarray}

In the subsequent two sections, we will consider general bright and dark soliton solutions for semi-discete coupled YO system (\ref{sdyo-20})-(\ref{sdyo-21}) in details.
For brevity, we refer to the soliton solution including bright or dark soliton for the SW components and bright one for the LW component as the bright or dark soliton solution.

\section{Bright soliton solution for the semi-discete coupled YO system}
In this section, we will construct bright soliton solution for the semi-discete coupled YO system (\ref{sdyo-20})-(\ref{sdyo-21}).
First, we briefly recall B\"{a}cklund transformations of the semi-discrete BKP hierarchy by the following Lemma \cite{ohta2009discretization}.

\textbf{Lemma 3.1}
The following bilinear equations
\begin{eqnarray}
\label{sdyo-22}&& D_t g^{(\mu)}_n \cdot f_n = g^{(\mu)}_{n+1}f_{n-1}-g^{(\mu)}_{n-1}f_{n+1},\\
\label{sdyo-23}&& D_t h^{(\mu)}_n \cdot f_n = h^{(\mu)}_{n+1}f_{n-1}-h^{(\mu)}_{n-1}f_{n+1},\\
\label{sdyo-24}&& D_{y^{(\mu)}} f_{n+1} \cdot f_n = g^{(\mu)}_{n+1} h^{(\mu)}_n- g^{(\mu)}_n h^{(\mu)}_{n+1},
\end{eqnarray}
for $\mu=1,\cdots,M$ are satisfied by the pfaffians
\begin{eqnarray}
\label{sdyo-25}&& f_n={\rm pf}(a_1,\cdots, a_{2N},c_{2N},\cdots,c_1),\\
\label{sdyo-26}&& g^{(\mu)}_n={\rm pf}(d_0,a_1,\cdots, a_{2N},c_{2N},\cdots,c_1,\alpha^{(\mu)}),\\
\label{sdyo-27}&& h^{(\mu)}_n={\rm pf}(d_0,a_1,\cdots, a_{2N},c_{2N},\cdots,c_1,\beta^{(\mu)}),
\end{eqnarray}
where the pfaffian elements are defined by
\begin{eqnarray*}
&& {\rm pf}(a_j,a_k)=\frac{p_j-p_k}{p_jp_k-1}(p_jp_k)^n\exp(\xi_j+\xi_k), \\
&& {\rm pf}(d_l,a_j)=p^{n+l}_j\exp(\xi_j),\ \ {\rm pf}(a_j,c_k)=\delta_{jk},\\
&&  {\rm pf}(d_l,c_j)={\rm pf}(d_l,\alpha^{(\mu)})={\rm pf}(d_l,\beta^{(\mu)})={\rm pf}(a_j,\alpha^{(\mu)})={\rm pf}(a_j,\beta^{(\mu)})=0,\\
&& {\rm pf}(c_j,c_k)=\left\{
              \begin{array}{ll}
                \sum^M_{\mu=1}\frac{\exp(\zeta^{(\mu)}_j+\eta^{(\mu)}_k)}{Q^{(\mu)}_j+P^{(\mu)}_k}, & 2N\geq j \geq N+1, N\geq k\geq 1 ; \\
                0, & \hbox{otherwise,}
              \end{array}
            \right.\\
&& {\rm pf}(c_j,\alpha^{(\mu)})=\left\{
              \begin{array}{ll}
                \exp(\eta^{(\mu)}_j), &  N\geq j\geq 1 ; \\
                0, & 2N\geq j \geq N+1,
              \end{array}
            \right.\\
&&  {\rm pf}(c_j,\beta^{(\mu)})=\left\{
              \begin{array}{ll}
                0, &  N\geq j\geq 1 ; \\
                 \exp(\zeta^{(\mu)}_j), & 2N\geq j \geq N+1,
              \end{array}
            \right.
\end{eqnarray*}
with
\begin{eqnarray*}
\xi_j=(p_j-\frac{1}{p_j})t,\ \
\zeta^{(\mu)}_j=Q^{(\mu)}_j y^{(\mu)}+ \zeta^{(\mu)}_{j0},\ \ \eta^{(\mu)}_j=P^{(\mu)}_j y^{(\mu)}+ \eta^{(\mu)}_{j0}.
\end{eqnarray*}
Here $p_j$, $Q^{(\mu)}_j$, $P^{(\mu)}_j$,$\zeta^{(\mu)}_{j0}$ and $\eta^{(\mu)}_{j0}$ are arbitrary constants.

\textbf{Proof}
From the definition of the functions $f_n$, $g^{(\mu)}_n$ and $h^{(\mu)}_n$, we can derive the following
pfaffian's rules:
\begin{eqnarray*}
&&\hspace{-1cm} f_{n+1}={\rm pf}(d_0,d_1,\bullet),\ \ f_{n-1}={\rm pf}(d_0,d_{-1},\bullet),\ \
 \partial_tf_{n}={\rm pf}(d_{-1},d_1,\bullet),\\
&&\hspace{-1cm} \partial_{y^{(\mu)}}f_{n}={\rm pf}(\bullet,\alpha^{(\mu)},\beta^{(\mu)}),\ \
 \partial_{y^{(\mu)}}f_{n+1}={\rm pf}(d_0,d_1,\bullet,\alpha^{(\mu)},\beta^{(\mu)}),\\
%%%%%%%%%%%%%%%%%%%%%%%%%%%%%%
&&\hspace{-1cm} g^{(\mu)}_{n+1}={\rm pf}(d_1,\bullet,\alpha^{(\mu)}),\ \
 g^{(\mu)}_{n-1}={\rm pf}(d_{-1},\bullet,\alpha^{(\mu)}),\ \
 \partial_tg^{(\mu)}_{n}={\rm pf}(d_0,d_{-1},d_1,\bullet,\alpha^{(\mu)}),\\
%%%%%%%%%%%%%%%%%%%%%%%%%%%%%%
&&\hspace{-1cm} h^{(\mu)}_{n+1}={\rm pf}(d_1,\bullet,\beta^{(\mu)}),\ \
 h^{(\mu)}_{n-1}={\rm pf}(d_{-1},\bullet,\beta^{(\mu)}),\ \
 \partial_th^{(\mu)}_{n}={\rm pf}(d_0,d_{-1},d_1,\bullet,\beta^{(\mu)}),
\end{eqnarray*}
where
\begin{eqnarray*}
{\rm pf}(d_0,d_1)={\rm pf}(d_0,d_{-1})=1,\ \ {\rm pf}(d_{-1},d_1)={\rm pf}(\alpha^{(\mu)},\beta^{(\mu)})=0,
\end{eqnarray*}
and $(\bullet)=(a_1,\cdots, a_{2N},c_{2N},\cdots,c_1)$.

Now the algebraic identities of Pfaffian
\begin{eqnarray*}
&&\hspace{-2cm} {\rm pf}(d_0,d_{-1},d_1,\bullet,\alpha^{(\mu)}) {\rm pf}(\bullet)
={\rm pf}(d_0,d_{-1},\bullet){\rm pf}(d_1,\bullet,\alpha^{(\mu)})\\
&&\hspace{2cm}-{\rm pf}(d_0,d_1,\bullet){\rm pf}(d_{-1},\bullet,\alpha^{(\mu)})
+{\rm pf}(d_0,\bullet,\alpha^{(\mu)}){\rm pf}(d_{-1},d_1,\bullet),\\
%%%%%%%%%%%%%%%%%%%%%%%%%%%%%%%%
&&\hspace{-2cm} {\rm pf}(d_0,d_{-1},d_1,\bullet,\beta^{(\mu)}) {\rm pf}(\bullet)
={\rm pf}(d_0,d_{-1},\bullet){\rm pf}(d_1,\bullet,\beta^{(\mu)})\\
&&\hspace{2cm}-{\rm pf}(d_0,d_1,\bullet){\rm pf}(d_{-1},\bullet,\beta^{(\mu)})
+{\rm pf}(d_0,\bullet,\beta^{(\mu)}){\rm pf}(d_{-1},d_1,\bullet),
\end{eqnarray*}
and
\begin{eqnarray*}
&&\hspace{-2cm} {\rm pf}(d_0,d_1,\bullet,\alpha^{(\mu)},\beta^{(\mu)}) {\rm pf}(\bullet)
={\rm pf}(d_0,d_{1},\bullet){\rm pf}(\bullet,\alpha^{(\mu)},\beta^{(\mu)})\\
&&\hspace{2cm}-{\rm pf}(d_0,\bullet,\alpha^{(\mu)}){\rm pf}(d_1,\bullet,\beta^{(\mu)})
+{\rm pf}(d_1,\bullet,\alpha^{(\mu)}){\rm pf}(d_0,\bullet,\beta^{(\mu)}),
\end{eqnarray*}
together with the above Pfaffian expressions of tau functions give the bilinear
equations ({\ref{sdyo-22}})-(\ref{sdyo-24}). $\square$

Assume $c=0$ in Eq.(\ref{sdyo-18}), which implies the bright-type soliton solution for semi-discete coupled YO system.
Through the dependent variable transformation
\begin{eqnarray}
\label{sdyo-28} S^{(\mu)}_n={\rm i}^n\frac{g^{(\mu)}_n}{f_n}, \ \ \bar{S}^{(\mu)}_n=(-{\rm i})^n\frac{\bar{g}^{(\mu)}_n}{f_n},\ \ L_n=2\left( \frac{f_{n+1}f_{n-1}}{f^2_n}-1 \right),\ \ \hat{\hat{L}}=2\ln f_n,
\end{eqnarray}
Eqs. ({\ref{sdyo-17}})-(\ref{sdyo-19}) are cast into
\begin{eqnarray}
\label{sdyo-29} && D_tg^{(\mu)}_n\bullet f_n =g^{(\mu)}_{n+1}f_{n-1}-g^{(\mu)}_{n-1}f_{n+1},\\
\label{sdyo-30} && D_t\bar{g}^{(\mu)}_n\bullet f_n =\bar{g}^{(\mu)}_{n+1}f_{n-1}-\bar{g}^{(\mu)}_{n-1}f_{n+1},\\
\label{sdyo-31} && D_t f_{n+1}\bullet f_n = \sum^{M}_{\mu=1}{\rm i}\frac{c^{(\mu)}}{2}(g^{(\mu)}_{n+1}\bar{g}^{(\mu)}_n-g^{(\mu)}_{n}\bar{g}^{(\mu)}_{n+1}) ,
\end{eqnarray}
for $\mu=1,\cdots,M$.

In order to carry out the dimension reduction, we define
\begin{eqnarray*}
&& {\rm pf}(a'_j,a'_k)=\frac{p_j-p_k}{p_jp_k-1}(p_jp_k)^n, \\
&& {\rm pf}(d_l,a'_j)=p^{n+l}_j,\ \ {\rm pf}(a'_j,c'_k)=\delta_{jk},\\
&&  {\rm pf}(d_l,c'_j)={\rm pf}(d_l,\alpha^{(\mu)})={\rm pf}(d_l,\beta^{(\mu)})={\rm pf}(a'_j,\alpha^{(\mu)})={\rm pf}(a'_j,\beta^{(\mu)})=0,\\
&& {\rm pf}(c'_j,c'_k)=\left\{
              \begin{array}{ll}
                \sum^M_{\mu=1}\frac{\exp(\zeta^{(\mu)}_j+\eta^{(\mu)}_k+\xi_j+\xi_k)}{Q^{(\mu)}_j+P^{(\mu)}_k}, & 2N\geq j \geq N+1, N\geq k\geq 1 ; \\
                0, & \hbox{otherwise,}
              \end{array}
            \right.\\
&& {\rm pf}(c'_j,\alpha^{(\mu)})=\left\{
              \begin{array}{ll}
                \exp(\eta^{(\mu)}_j+\xi_j), &  N\geq j\geq 1 ; \\
                0, & 2N\geq j \geq N+1,
              \end{array}
            \right.\\
&&  {\rm pf}(c'_j,\beta^{(\mu)})=\left\{
              \begin{array}{ll}
                0, &  N\geq j\geq 1 ; \\
                 \exp(\zeta^{(\mu)}_j+\xi_j), & 2N\geq j \geq N+1.
              \end{array}
            \right.
\end{eqnarray*}
Then, the Pfaffians $f_n$, $g_n^{(\mu)}$ and $h_n^{(\mu)}$ in Eqs.(\ref{sdyo-25})-(\ref{sdyo-27})
have alternative expressions in pfaffians
\begin{eqnarray}
\label{sdyo-32}&& f_n={\rm pf}(a'_1,\cdots, a'_{2N},c'_{2N},\cdots,c'_1),\\
\label{sdyo-33}&& g^{(\mu)}_n={\rm pf}(d_0,a'_1,\cdots, a'_{2N},c'_{2N},\cdots,c'_1,\alpha^{(\mu)}),\\
\label{sdyo-34}&& h^{(\mu)}_n={\rm pf}(d_0,a'_1,\cdots, a'_{2N},c'_{2N},\cdots,c'_1,\beta^{(\mu)}).
\end{eqnarray}

Therefore, under the reduction conditions,
\begin{eqnarray}
\label{sdyo-35}&&\hspace{-1cm} P^{(\mu)}_j=\frac{1}{s^{(\mu)}}(p_j-\frac{1}{p_j}),\ \ N\geq j\geq 1;\ \
 Q^{(\mu)}_j=\frac{1}{s^{(\mu)}}(p_j-\frac{1}{p_j}),\ \ 2N\geq j \geq N+1,
\end{eqnarray}
the following relation holds
\begin{eqnarray}
\label{sdyo-36} \partial_t(f_n,g^{(\nu)}_n,h^{(\nu)}_n)=\sum^M_{\mu=1}s^{(\mu)}\partial_{y^{(\mu)}}(f_n,g^{(\nu)}_n,h^{(\nu)}_n),
\end{eqnarray}
and thus one can get
\begin{eqnarray}
\label{sdyo-37} D_t f_{n+1} \cdot f_n = \sum^M_{\mu=1}s^{(\mu)}(g^{(\mu)}_{n+1} h^{(\mu)}_n- g^{(\mu)}_n h^{(\mu)}_{n+1}).
\end{eqnarray}

Lastly, by taking the complex conjugate conditions
\begin{eqnarray}
\label{sdyo-38} s^{(\mu)}={\rm i}\frac{c^{(\mu)}}{2},\ \ p_{2N+1-j}=\bar{p}_j,\ \ \zeta^{(\mu)}_{p_{2N+1-j,0}}=\bar{\eta}^{(\mu)}_{j,0},\ \  \mbox{for} \ \  N\geq j\geq 1,
\end{eqnarray}
and $y^{(\mu)}$ are pure imaginary, then the function $h^{(\mu)}_n=\bar{g}^{(\mu)}_n$, and Eqs.({\ref{sdyo-22}}), (\ref{sdyo-23}) and (\ref{sdyo-37}) become Eqs.({\ref{sdyo-29}})--(\ref{sdyo-31}).

To summarize, we arrive at a Theorem below:

\textbf{Theorem 3.1}
The bilinear equations ({\ref{sdyo-29}})--(\ref{sdyo-31}) are satisfied by the Pfaffians,
\begin{eqnarray}
\label{sdyo-39} && f_n={\rm pf}(a'_1,\cdots, a'_{2N},c'_{2N},\cdots,c'_1),\\
\label{sdyo-40} && g^{(\mu)}_n={\rm pf}(d_0,a'_1,\cdots, a'_{2N},c'_{2N},\cdots,c'_1,\alpha^{(\mu)}),\\
\label{sdyo-41} && \bar{g}^{(\mu)}_n={\rm pf}(d_0,a'_1,\cdots, a'_{2N},c'_{2N},\cdots,c'_1,\beta^{(\mu)}),
\end{eqnarray}
where the pfaffian elements are defined by
\begin{eqnarray*}
&& {\rm pf}(a'_j,a'_k)=\frac{p_j-p_k}{p_jp_k-1}(p_jp_k)^n, \\
&& {\rm pf}(d_l,a'_j)=p^{n+l}_j,\ \ {\rm pf}(a'_j,c'_k)=\delta_{jk},\\
&&  {\rm pf}(d_l,c'_j)={\rm pf}(d_l,\alpha^{(\mu)})={\rm pf}(d_l,\beta^{(\mu)})={\rm pf}(a'_j,\alpha^{(\mu)})={\rm pf}(a'_j,\beta^{(\mu)})=0,\\
&& {\rm pf}(c'_j,c'_k)=\left\{
              \begin{array}{ll}
                \sum^M_{\mu=1}\frac{\exp(\zeta^{(\mu)}_j+\eta^{(\mu)}_k+\xi_j+\xi_k)}{Q^{(\mu)}_j+P^{(\mu)}_k}, & 2N\geq j \geq N+1, N\geq k\geq 1 ; \\
                0, & \hbox{otherwise,}
              \end{array}
            \right.\\
&& {\rm pf}(c'_j,\alpha^{(\mu)})=\left\{
              \begin{array}{ll}
                \exp(\eta^{(\mu)}_j+\xi_j), &  N\geq j\geq 1 ; \\
                0, & 2N\geq j \geq N+1,
              \end{array}
            \right.\\
&&  {\rm pf}(c'_j,\beta^{(\mu)})=\left\{
              \begin{array}{ll}
                0, &  N\geq j\geq 1 ; \\
                 \exp(\zeta^{(\mu)}_j+\xi_j), & 2N\geq j \geq N+1.
              \end{array}
            \right.
\end{eqnarray*}
with
\begin{eqnarray*}
&& P^{(\mu)}_j=-{\rm i}\frac{2}{c^{(\mu)}}(p_j-\frac{1}{p_j}) \ \ \mbox{for} \ \   N\geq j\geq 1; \\
&& Q^{(\mu)}_j=-{\rm i}\frac{2}{c^{(\mu)}}(p_j-\frac{1}{p_j})\ \ \mbox{for} \ \ 2N\geq j \geq N+1,\\
&& \xi_j=(p_j-\frac{1}{p_j})t,\ \
\zeta^{(\mu)}_j= \zeta^{(\mu)}_{j,0},\ \ \eta^{(\mu)}_j=\eta^{(\mu)}_{j,0},\\
\end{eqnarray*}
and $p_j$, $\zeta^{(\mu)}_j\equiv \zeta^{(\mu)}_{j,0}$ and $\eta^{(\mu)}_j\equiv\eta^{(\mu)}_{j,0}$ are constants satisfying
 $p_{2N+1-j}=\bar{p}_j$, $\zeta^{(\mu)}_{2N+1-j,0}=\bar{\eta}^{(\mu)}_{j,0}$ for $N\geq j\geq 1$.

%On page 8, in the explanation of the Pfaffian elements:
% We don't need $y^{(\mu)}$ or $z^{(\mu)}$, and we can fix $y^{(\mu)}=0$.
% Thus we can remove the line of $\zeta_j^{(\mu)}=Q_j...$ and
% $\eta_j^{(\mu)}=P_j ...$.
% Also we can just replace $\zeta_j^{(\mu)}$ and $\eta_j^{(\mu)}$
% in $pf(c_j',c_k')$, $pf(c_j',\alpha^{(\mu)}$, $pf(c_j',\beta^{(\mu)}$
% by $\zeta_{j0}^{(\mu)}$ and $\eta_{j0}^{(\mu)}$.

In what follows, we will illustrate one and two bright soliton solutions for $M=2$.

\textbf{One-soliton solution:} By taking $N=1$ in ({\ref{sdyo-39}})-(\ref{sdyo-41}),
we get the tau functions for the one-soliton solution
\begin{eqnarray}
\label{sdyo-42} && f_n=1+\frac{{\rm i}}{2}\frac{(c^{(1)}A_1\bar{A}_1+c^{(2)}B_1\bar{B}_1)p_1\bar{p}_1}{(p_1\bar{p}_1-1)^2}\frac{p_1-\bar{p}_1}{p_1+\bar{p}_1}(p_1\bar{p}_1)^n\exp(\xi_1+\bar{\xi}_1),\\
\label{sdyo-43} && g^{(1)}_n=A_1p^n_1\exp(\xi_1),\ \  g^{(2)}_n=B_1p^n_1\exp(\xi_1),
\end{eqnarray}
where $\xi_1= (p_1-\frac{1}{p_1})t$, $A_1=\exp[\eta^{(1)}_{1,0}]$ and $B_1=\exp[\eta^{(2)}_{1,0}]$.

In order to avoid the singularity, the condition $\frac{{\rm i} (c^{(1)}A_1\bar{A}_1+c^{(2)}B_1\bar{B}_1)(p_1-\bar{p}_1)}{2 (p_1+\bar{p}_1)}>0$ need to be satisfied. Further, the above tau functions lead to the one-soliton solution as follows
\begin{eqnarray}
&& S^{(1)}_n=\frac{A_1}{2}  \exp({\rm i}\xi'_{1I}-\theta_0) {\rm sech}(\xi'_{1R}+\theta_0),\\
&& S^{(2)}_n=\frac{B_1}{2}  \exp({\rm i}\xi'_{1I}-\theta_0) {\rm sech}(\xi'_{1R}+\theta_0),\\
&& L_n= \frac{(p_1\bar{p}_1-1)^2}{2p_1\bar{p}_1} {\rm sech}^2(\xi'_{1R}+\theta_0),
\end{eqnarray}
where $\xi'_1=\xi'_{1R}+{\rm i}\xi'_{1I}=n\ln({\rm i}p_1)+\xi_1$ and $\exp(2\theta_0)=\frac{{\rm i}}{2}\frac{(c^{(1)}A_1\bar{A}_1+c^{(2)}B_1\bar{B}_1)p_1\bar{p}_1}{(p_1\bar{p}_1-1)^2}\frac{p_1-\bar{p}_1}{p_1+\bar{p}_1}$.
The quantities $\frac{|A|}{2}\exp(-\theta_0)$ and $\frac{|B|}{2}\exp(-\theta_0)$
represent the amplitudes of the bright
solitons in the SW components $S^{(1)}_n$ and $S^{(2)}_n$ respectively.
The real quantity $\frac{(p_1\bar{p}_1-1)^2}{2p_1\bar{p}_1}$ denotes the amplitude of soliton in the LW component.
As an example, we illustrate one-soliton in Figure 1 for the nonlinearity coefficients $(c^{(1)},c^{(2)})=(1,-1)$.
The parameters are chosen as $A_1=1$, $B_1=1+{\rm i}$ and $p_1=1+{\rm i}$.

\begin{figure}[!htbp]
\centering
{\includegraphics[height=1.8in,width=2.4in]{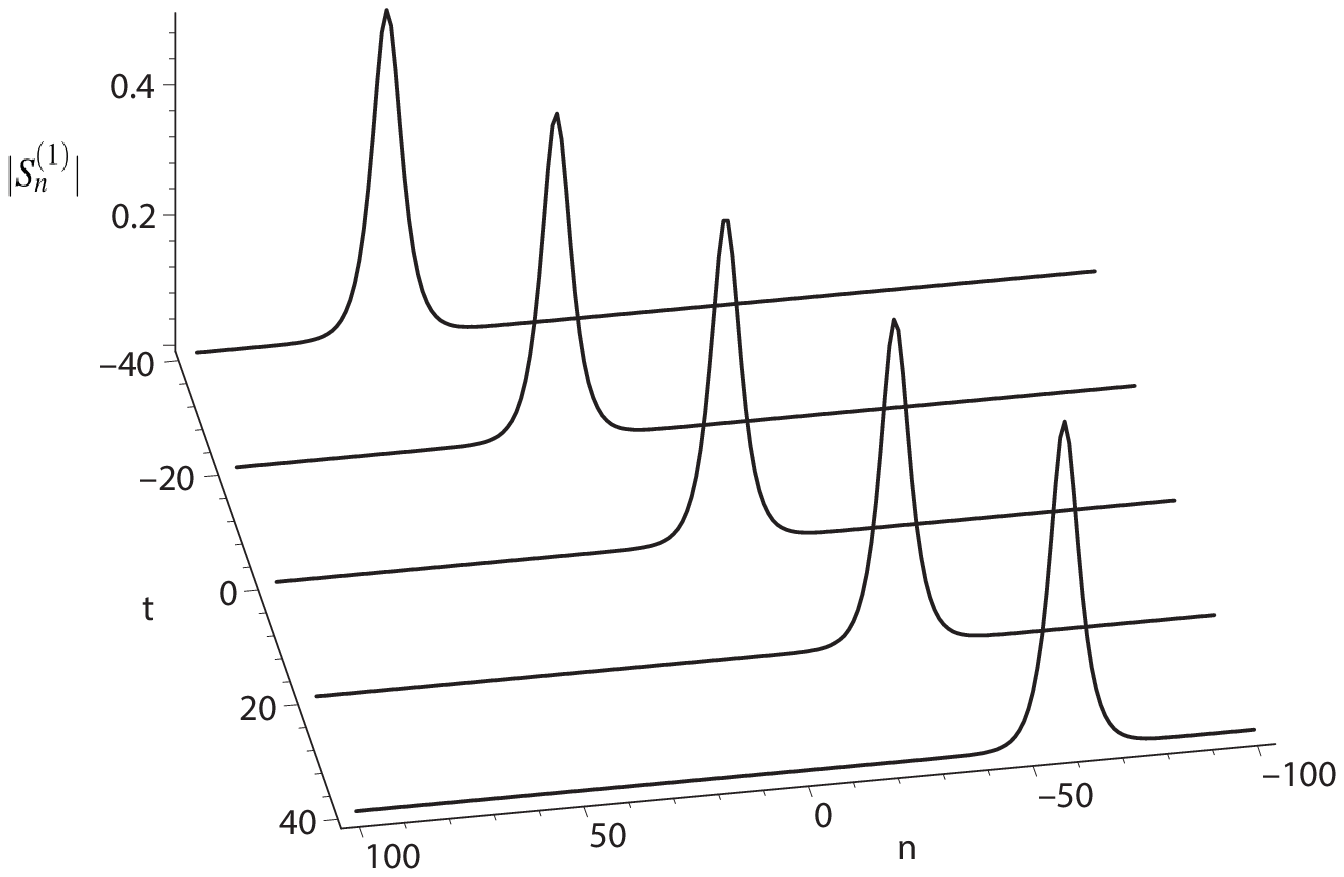}}
{\includegraphics[height=1.8in,width=2.4in]{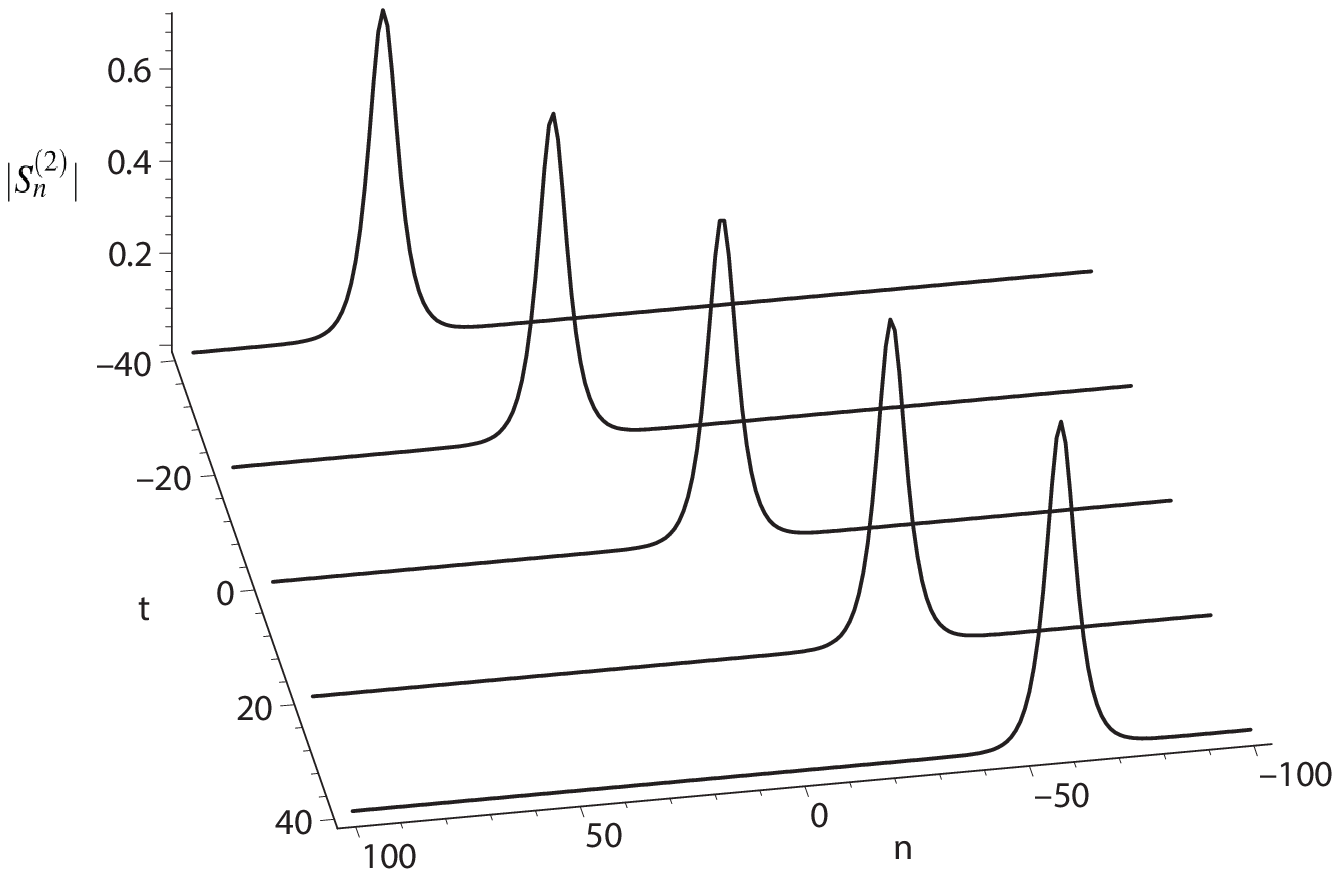}}
{\includegraphics[height=1.8in,width=2.4in]{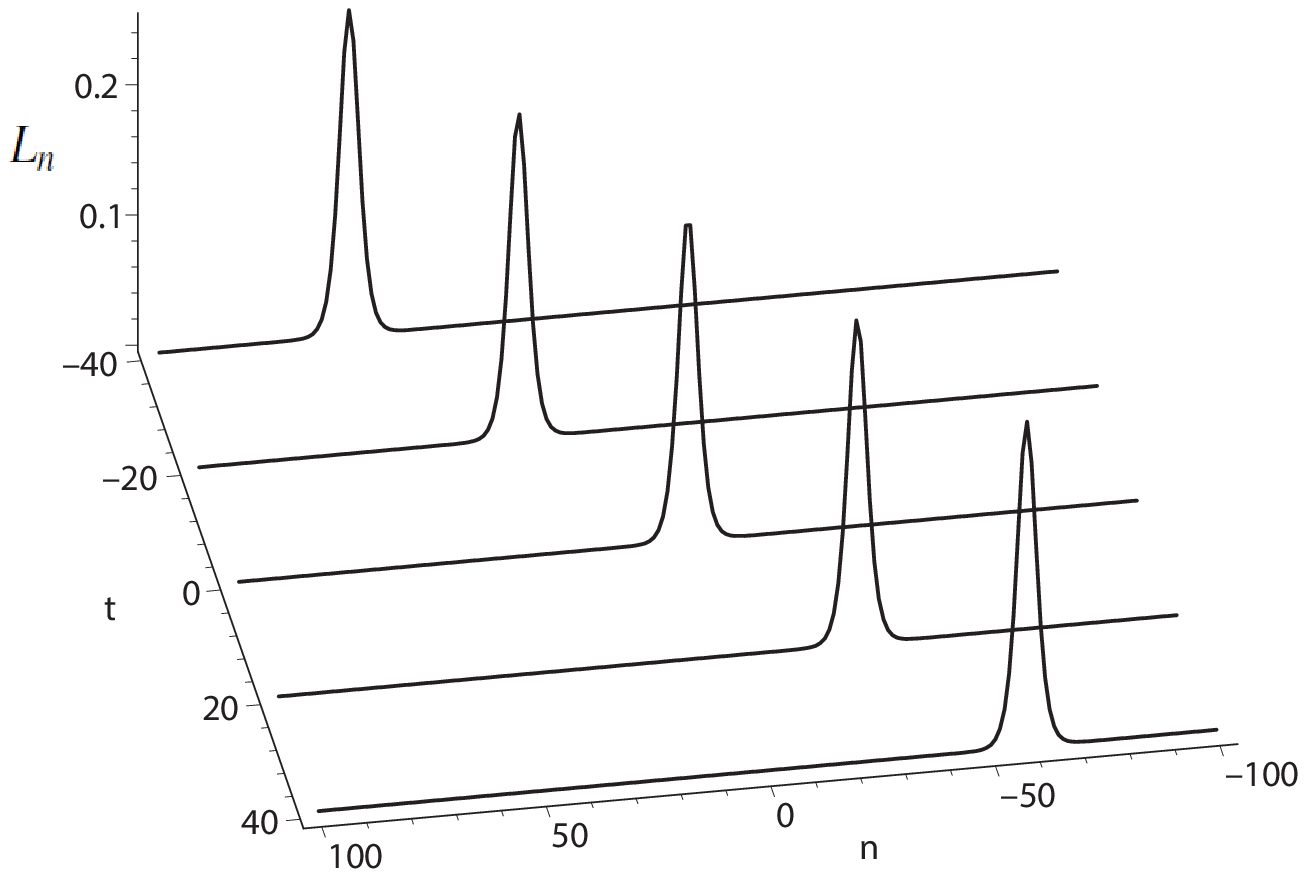}}
\caption{ The profiles of evolutions of one-soliton solution (bright soliton for SW components).
%with the parameters $
%(c^{(1)},c^{(2)})=(1,-1)$, $A_1=1$, $B_1=1+{\rm i}$ and $p_1=1+{\rm i}$.
}
\end{figure}

\textbf{Two-soliton solution:} By taking $N=2$ in ({\ref{sdyo-39}})-(\ref{sdyo-41}),
we get the tau functions for the two-soliton solution
\begin{eqnarray}
\nonumber f_n&=&1+C_{1\bar{1}}E_1\bar{E}_1+C_{1\bar{2}}E_1\bar{E}_2+C_{2\bar{1}}E_2\bar{E}_1+C_{2\bar{2}}E_2\bar{E}_2,\\
\label{sdyo-44} && +|P_{12}|^2(P_{1\bar{1}}P_{2\bar{2}}C_{1\bar{2}}C_{2\bar{1}}-P_{1\bar{2}}P_{2\bar{1}}C_{1\bar{1}}C_{2\bar{2}})E_1E_2\bar{E}_1\bar{E}_2,\\
\nonumber g^{(1)}_n&=&A_1E_1+A_2E_2+P_{12}(A_1P_{1\bar{1}}C_{2\bar{1}}-A_2P_{2\bar{1}}C_{1\bar{1}})E_1E_2\bar{E}_1\\
\label{sdyo-45}&&+P_{12}(A_1P_{1\bar{2}}C_{2\bar{2}}-A_2P_{2\bar{2}}C_{1\bar{2}})E_1E_2\bar{E}_2,\\
\nonumber g^{(2)}_n&=&B_1E_1+B_2E_2+P_{12}(B_1P_{1\bar{1}}C_{2\bar{1}}-B_2P_{2\bar{1}}C_{1\bar{1}})E_1E_2\bar{E}_1\\
\label{sdyo-46}&&+P_{12}(B_1P_{1\bar{2}}C_{2\bar{2}}-B_2P_{2\bar{2}}C_{1\bar{2}})E_1E_2\bar{E}_2,
\end{eqnarray}
with
\begin{eqnarray*}
&& P_{12}=\frac{p_1-p_2}{p_1p_2-1},\ \ P_{j\bar{k}}=\frac{p_j-\bar{p}_k}{p_j\bar{p}_k-1},\\
&& E_j=p^n_j\exp(\xi_j),\ \ C_{j\bar{k}}=\frac{{\rm i}}{2}\frac{(c^{(1)}A_j\bar{A}_k+c^{(2)}B_j\bar{B}_k)p_j\bar{p}_k}{(p_j\bar{p}_k-1)^2}\frac{p_j-\bar{p}_k}{p_j+\bar{p}_k},
\end{eqnarray*}
where $\xi_j= (p_j-\frac{1}{p_j})t$, $A_j=\exp[\eta^{(1)}_{j,0}]$ and $B_j=\exp[\eta^{(2)}_{j,0}]$ for $j=1,2$.
Figure 2 shows two-soliton interaction with the parameters as $
(c^{(1)},c^{(2)})=(1,-1)$, $A_1=2$, $A_2=5$, $B_1=3$, $B_2=4$, $p_1=1+{\rm i}$ and $p_2=-\frac{1}{2}+\frac{2}{3}{\rm i}$.

\begin{figure}[!htbp]
\centering
{\includegraphics[height=1.8in,width=2.4in]{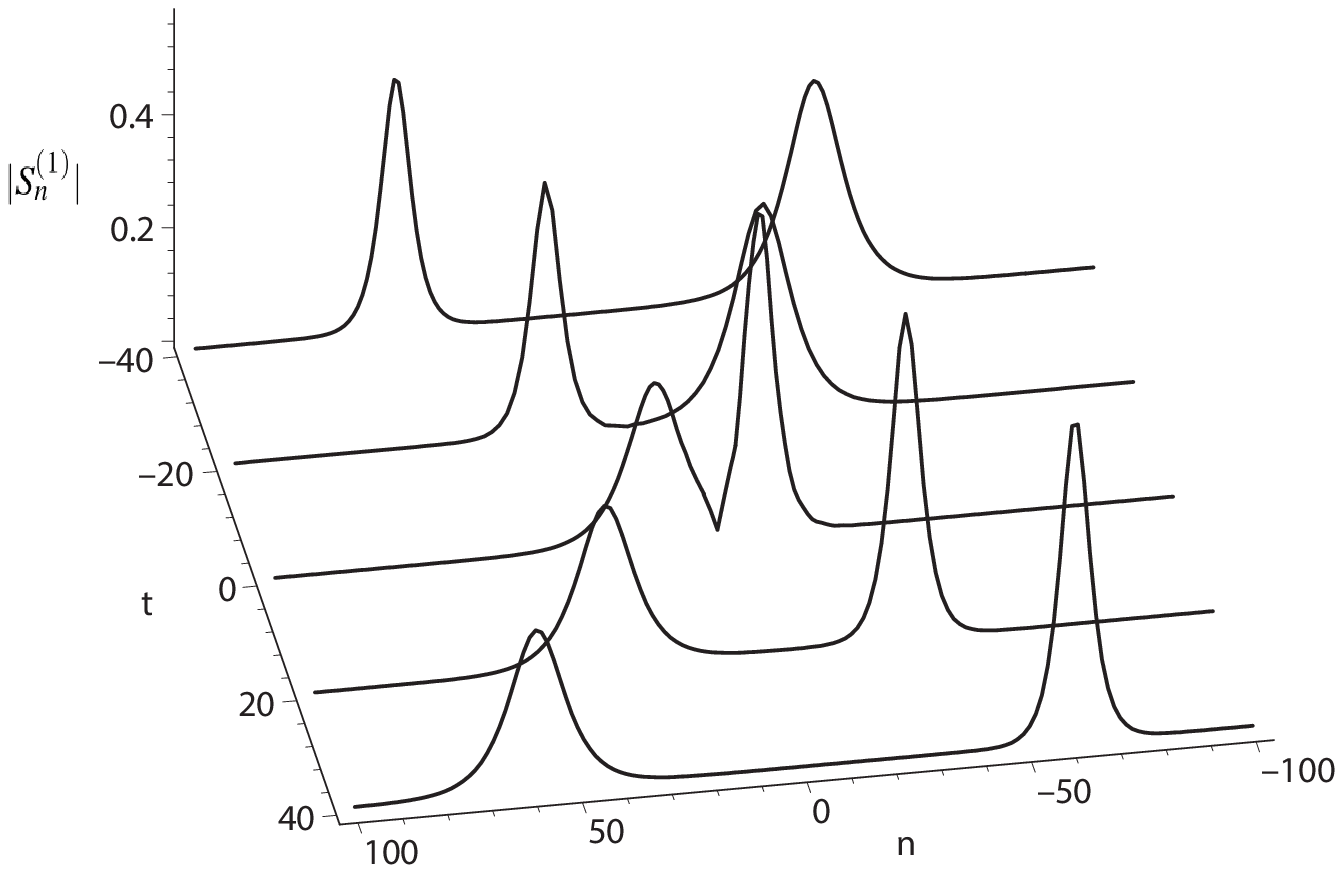}}
{\includegraphics[height=1.8in,width=2.4in]{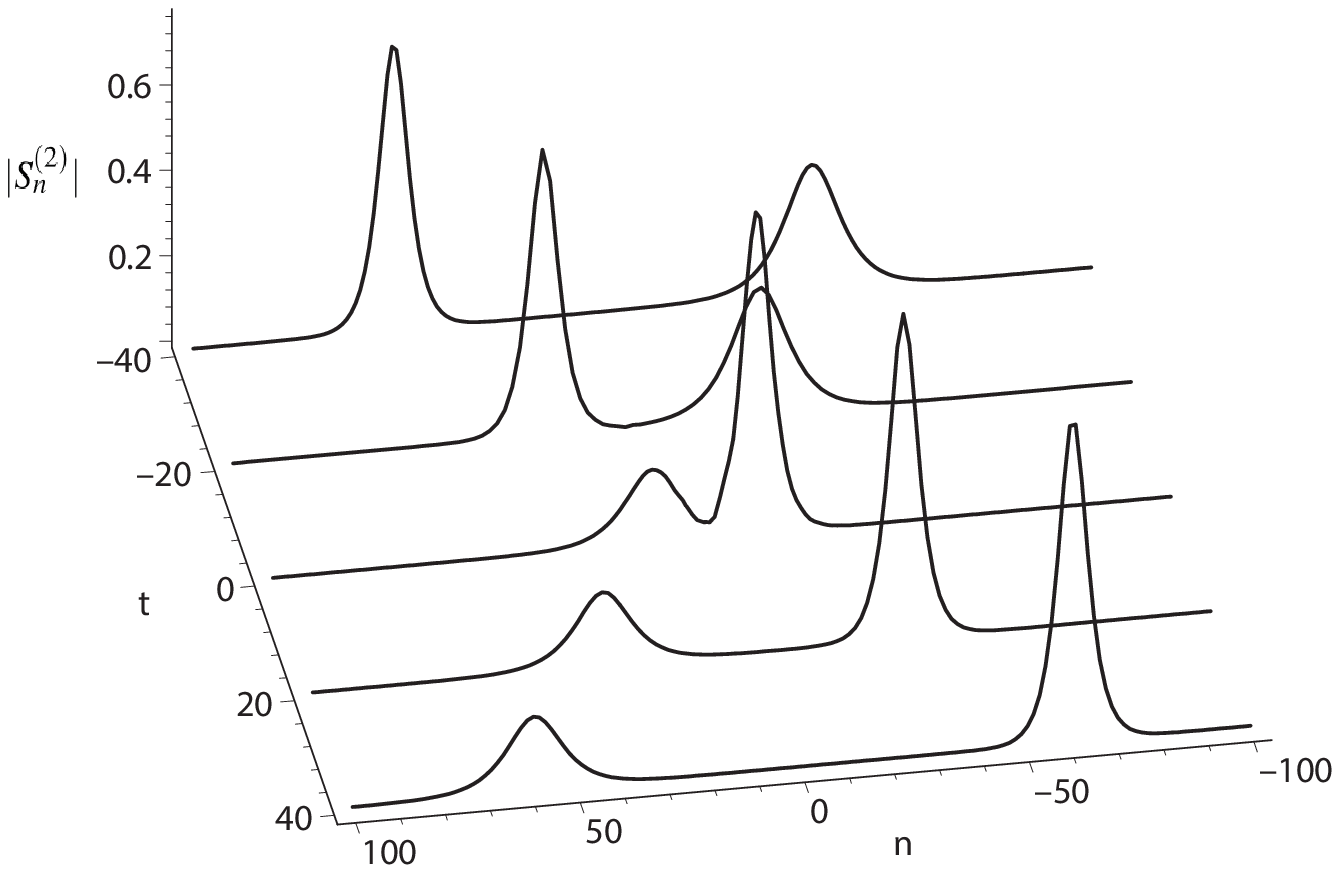}}
{\includegraphics[height=1.8in,width=2.4in]{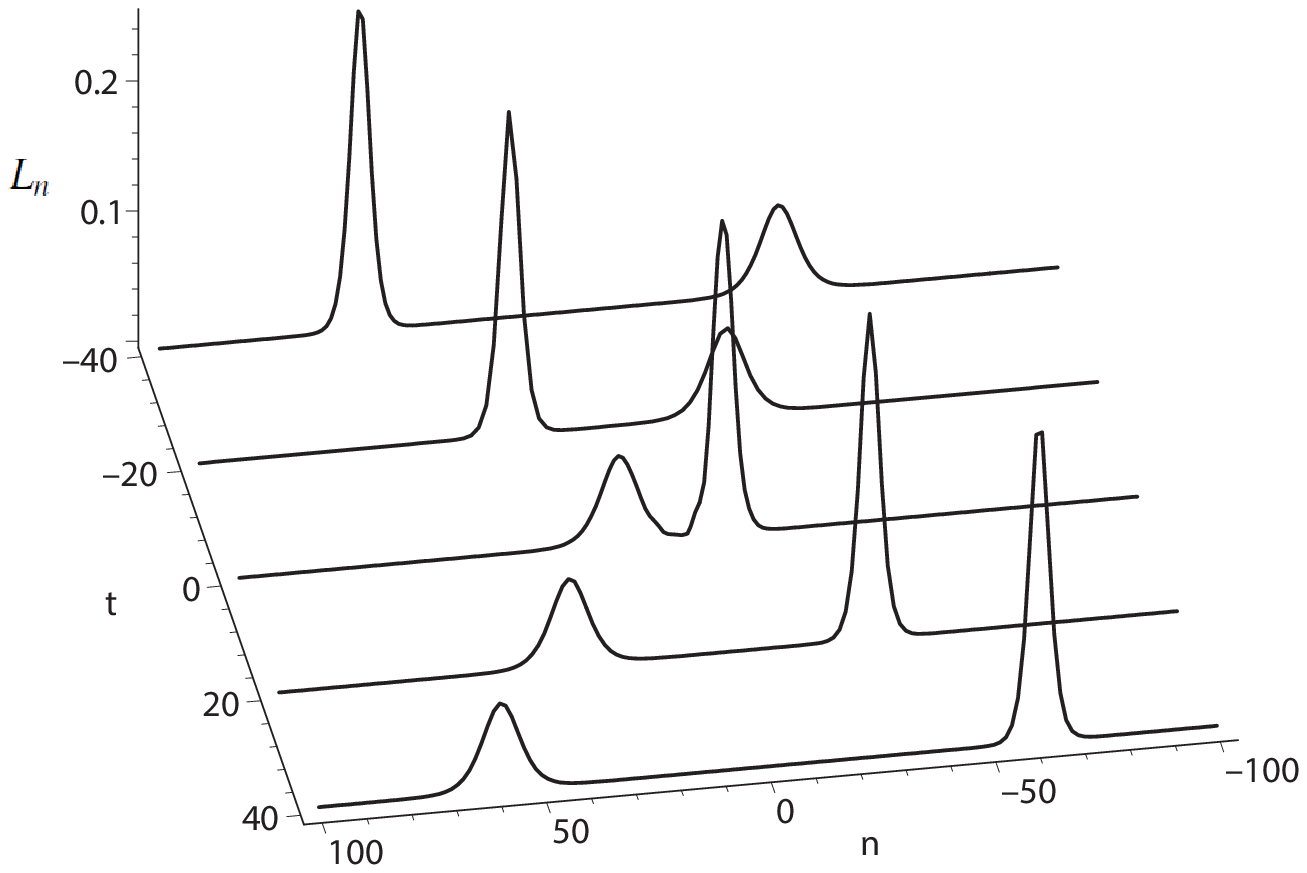}}
\caption{The profiles of evolutions of two-soliton solution (bright soliton for SW components).
%with the parameters $
%(c^{(1)},c^{(2)})=(1,-1)$, $A_1=2$, $A_2=5$, $B_1=3$, $B_2=4$ and $p_1=1+{\rm i}$, $p_2=-\frac{1}{2}+\frac{2}{3}{\rm i}$.
}
\end{figure}

\section{Dark soliton solution for the semi-discete coupled YO system}
In this section, we will consider dark soliton solution for the semi-discete coupled YO system (\ref{sdyo-20})-(\ref{sdyo-21}).
To this end, we need to introduce another set of B\"{a}cklund transformations of the semi-discrete BKP hierarchy by the following Lemma \cite{ohta2009discretization}.

\textbf{Lemma 4.1}
The following bilinear equations
\begin{eqnarray}
\label{sdyo-47}&& (D_t + \alpha^{(\mu)}-\frac{1}{\alpha^{(\mu)}} )g^{(\mu)}_n \cdot f_n = \alpha^{(\mu)}g^{(\mu)}_{n+1}f_{n-1}-\frac{1}{\alpha^{(\mu)}}g^{(\mu)}_{n-1}f_{n+1},\\
\label{sdyo-48}&& (D_t + \frac{1}{\alpha^{(\mu)}}-\alpha^{(\mu)} ) h^{(\mu)}_n \cdot f_n = \frac{1}{\alpha^{(\mu)}}h^{(\mu)}_{n+1}f_{n-1}-\alpha^{(\mu)}h^{(\mu)}_{n-1}f_{n+1},\\
\label{sdyo-49}&& (D_{y^{(\mu)}} -\alpha^{(\mu)}+\frac{1}{\alpha^{(\mu)}})f_{n+1} \cdot f_n = -\alpha^{(\mu)}g^{(\mu)}_{n+1} h^{(\mu)}_n +\frac{1}{\alpha^{(\mu)}} g^{(\mu)}_n h^{(\mu)}_{n+1},
\end{eqnarray}
for $\mu=1,\cdots,M$ are satisfied by the Pfaffians
\begin{eqnarray}
\label{sdyo-50} && f_n=\tau^{0\cdots0}_n,\ \
 g_n^{(\mu )}  = \tau _n^{0...\mathop {\mathop {1} \limits^{\smile}}\limits^{\mu} ...0},\ \
 h_n^{(\mu )}  = \tau _n^{0...\mathop {\mathop {-1} \limits^{\smile}}\limits^{\mu} ...0},
\end{eqnarray}
where the tau function $\tau^{l^{(1)}\cdots l^{(M)}}_n$ is the Pfaffian $\tau^{l^{(1)}\cdots l^{(M)}}_n={\rm pf}(a_1,a_2,\cdots,a_{2N})$,
whose elements are defined by
\begin{eqnarray*}
&& {\rm pf}(a_j,a_k)=c_{jk}+\frac{p_j-p_k}{p_jp_k-1}{\rm pf}(d_0,a_j){\rm pf}(d_0,a_k),\\
&& {\rm pf}(d_l,a_j)=p^{n+l}_j\prod^M_{\mu=1}\left( \frac{p_j-\alpha^{(\mu)}}{1-\alpha^{(\mu)}p_j} \right)^{l^{(\mu)}}\exp(\xi_j),\\
&& \xi_j=(p_j-\frac{1}{p_j})t + \sum^M_{\mu=1}\left( \frac{p_j-\alpha^{(\mu)}}{1-\alpha^{(\mu)}p_j}-\frac{1-\alpha^{(\mu)}p_j}{p_j-\alpha^{(\mu)}} \right)y^{(\mu)}+\xi_{j,0}.
\end{eqnarray*}

\textbf{Proof}
From the definition of the tau functions, we have the following formulae of
pfaffians:
\begin{eqnarray*}
&& \tau^{l^{(1)}\cdots l^{(M)}}_{n+1}={\rm pf}(d_0,d_1,\bullet),\ \
\tau^{l^{(1)}\cdots l^{(M)}}_{n-1}={\rm pf}(d_0,d_{-1},\bullet),\\
&& \partial_t\tau^{l^{(1)}\cdots l^{(M)}}_{n}={\rm pf}(d_{-1},d_{1},\bullet),\ \
\partial_{y^{(\mu)}}\tau^{l^{(1)}\cdots l^{(M)}}_{n}={\rm pf}(d^{(\mu)}_{-1},d^{(\mu)}_{1},\bullet),\\
&& (\partial_{y^{(\mu)}}-\alpha^{(\mu)}+\frac{1}{\alpha^{(\mu)}})\tau^{l^{(1)}\cdots l^{(M)}}_{n+1}={\rm pf}(d_0,d_1,d^{(\mu)}_{-1},d^{(\mu)}_{1},\bullet),\\
%%%%%%%%%%%%%%%%%%%%%%%%%%%%%%%%%%%%%%%%%%%%%%%%%%
&&  \tau^{l^{(1)}\cdots l^{(\mu)}+1\cdots l^{(M)}}_{n}={\rm pf}(d_0,d^{(\mu)}_1,\bullet),\\
&&  \alpha^{(\mu)}\tau^{l^{(1)}\cdots l^{(\mu)}+1\cdots l^{(M)}}_{n+1}={\rm pf}(d_1,d^{(\mu)}_1,\bullet),\\
&& \frac{1}{\alpha^{(\mu)}}\tau^{l^{(1)}\cdots l^{(\mu)}+1\cdots l^{(M)}}_{n-1}={\rm pf}(d_{-1},d^{(\mu)}_1,\bullet),\\
&& (\partial_t+\alpha^{(\mu)}-\frac{1}{\alpha^{(\mu)}})\tau^{l^{(1)}\cdots l^{(\mu)}+1\cdots l^{(M)}}_{n}={\rm pf}(d_0,d_{-1},d_1,d^{(\mu)}_1,\bullet),\\
%%%%%%%%%%%%%%%%%%%%%%%%%%%%%%%%%%%%%%%%%%%%%%%
&&  \tau^{l^{(1)}\cdots l^{(\mu)}-1\cdots l^{(M)}}_{n}={\rm pf}(d_0,d^{(\mu)}_{-1},\bullet),\\
%&&  \alpha^{(\mu)}\tau^{l^{(1)}\cdots l^{(\mu)}-1\cdots l^{(M)}}_{n-1}={\rm pf}(d_{-1},d^{(\mu)}_{-1},\bullet),\\
&&  \frac{1}{\alpha^{(\mu)}} \tau^{l^{(1)}\cdots l^{(\mu)}-1\cdots l^{(M)}}_{n+1}={\rm pf}(d_1,d^{(\mu)}_{-1},\bullet),
%&&  (\partial_t+\frac{1}{\alpha^{(\mu)}}-\alpha^{(\mu)})\tau^{l^{(1)}\cdots l^{(\mu)}-1\cdots l^{(M)}}_{n}={\rm pf}(d_0,d_{-1},d_{1},d^{(\mu)}_{-1},\bullet),
\end{eqnarray*}
where
\begin{eqnarray*}
&&{\rm pf}(d_0,d_1)={\rm pf}(d_0,d_{-1})=1,\ \ {\rm pf}(d_{-1},d_1)=0,\\
&& {\rm pf}(d^{(\mu)}_{-1},d^{(\mu)}_1)=0,\ \ {\rm pf}(d_0,d^{(\mu)}_1)={\rm pf}(d_0,d^{(\mu)}_{-1})=1,\\
&& {\rm pf}(d_1,d^{(\mu)}_1)=\alpha^{(\mu)},\ \ {\rm pf}(d_1,d^{(\mu)}_{-1})={\rm pf}(d_{-1},d^{(\mu)}_1)=\frac{1}{\alpha^{(\mu)}},\\
&& {\rm pf}(d^{(\mu)}_l,a_j)=p^n_j \left( \frac{p_j-\alpha^{(\mu)}}{1-\alpha^{(\mu)}p_j}\right)^l \prod^M_{\nu=1} \left( \frac{p_j-\alpha^{(\nu)}}{1-\alpha^{(\nu)}p_j}\right)^{l^{(\nu)}}\exp(\xi_j),
\end{eqnarray*}
and $(\bullet)=(a_1,\cdots, a_{2N})$.

Now the algebraic identity of pfaffians together with the previous rules for tau functions in pfaffians
\begin{eqnarray*}
&&\hspace{-2cm} {\rm pf}(d_0,d_{-1},d_1,d^{(\mu)}_1,\bullet) {\rm pf}(\bullet)
={\rm pf}(d_0,d_{-1},\bullet){\rm pf}(d_1,d_1^{(\mu)},\bullet)\\
&&\hspace{2cm}-{\rm pf}(d_0,d_1,\bullet){\rm pf}(d_{-1},d_1^{(\mu)},\bullet)
+{\rm pf}(d_0,d_1^{(\mu)},\bullet){\rm pf}(d_{-1},d_1,\bullet),
\end{eqnarray*}
gives the bilinear equations (\ref{sdyo-47})-(\ref{sdyo-48}) while the algebraic identities of pfaffians
\begin{eqnarray*}
&&\hspace{-2cm} {\rm pf}(d_0,d_1,d_{-1}^{(\mu)},d_1^{(\mu)},\bullet) {\rm pf}(\bullet)
={\rm pf}(d_0,d_{1},\bullet){\rm pf}(d_{-1}^{(\mu)},d_1^{(\mu)},\bullet)\\
&&\hspace{2cm}-{\rm pf}(d_0,d_{-1}^{(\mu)},\bullet){\rm pf}(d_1,d_{1}^{(\mu)},\bullet)
+{\rm pf}(d_0,d_1^{(\mu)},\bullet){\rm pf}(d_1,d_{-1}^{(\mu)},\bullet),
\end{eqnarray*}
leads to the bilinear equation (\ref{sdyo-49}).$\square$

Next, we consider $c\neq0$ in (\ref{sdyo-18}), which implies the dark soliton solution of the semi-discrete YO system.
By applying the dependent variable transformation,
\begin{eqnarray}
\label{sdyo-51}&&\nonumber\hspace{-1.5cm} S^{(\mu)}_n=({\rm i}\alpha^{(\mu)})^n\exp[(\alpha^{(\mu)}-\bar{\alpha}^{(\mu)})t]\frac{g^{(\mu)}_n}{f_n}, \ \
\bar{S}^{(\mu)}_n=(-{\rm i}\bar{\alpha}^{(\mu)})^n\exp[(\bar{\alpha}^{(\mu)}-\alpha^{(\mu)})t]\frac{\bar{g}^{(\mu)}_n}{f_n},\\
&&\hspace{-1.5cm} L_n=2\left( \frac{f_{n+1}f_{n-1}}{f^2_n}-1 \right),\ \ \hat{\hat{L}}=2\ln f_n
\end{eqnarray}
with $|\alpha^{(\mu)}|=1$, Eqs. ({\ref{sdyo-17}})-(\ref{sdyo-19}) are cast into
\begin{eqnarray}
\label{sdyo-52}&& (D_t+\alpha^{(\mu)}-\bar{\alpha}^{(\mu)})g^{(\mu)}_n\bullet f_n =\alpha^{(\mu)} g^{(\mu)}_{n+1}f_{n-1}-\bar{\alpha}^{(\mu)}g^{(\mu)}_{n-1}f_{n+1},\\
\label{sdyo-53}&& (D_t+\bar{\alpha}^{(\mu)}-\alpha^{(\mu)})\bar{g}^{(\mu)}_n\bullet f_n =\bar{\alpha}^{(\mu)} \bar{g}^{(\mu)}_{n+1}f_{n-1}-\alpha^{(\mu)}\bar{g}^{(\mu)}_{n-1}f_{n+1},\\
\label{sdyo-54}&& D_t f_{n+1}\bullet f_n -c f_{n+1}\bullet f_n = \sum^M_{\mu=1}{\rm i}\frac{c^{(\mu)}}{2}(\alpha^{(\mu)} g^{(\mu)}_{n+1}\bar{g}^{(\mu)}_n-\bar{\alpha}^{(\mu)}g^{(\mu)}_{n}\bar{g}^{(\mu)}_{n+1}) ,
\end{eqnarray}
for  $\mu=1,\cdots,M$.

Now we carry out the reductions to obtain bilinear equations ({\ref{sdyo-52}})-(\ref{sdyo-54}) from Eqs.(
\ref{sdyo-47})-(\ref{sdyo-49}) in \textbf{Lemma 4.1}.
First, by taking
\begin{eqnarray}\label{sdyo-55}
c_{jk}=\delta_{2N+1-j,k},\ \ j<k,
\end{eqnarray}
the tau functions $\tau^{l^{(1)}\cdots l^{(M)}}_n$ can be rewritten as
\begin{eqnarray}\label{sdyo-56}
\tau^{l^{(1)}\cdots l^{(M)}}_n= {\rm pf}(a'_1,a'_2,\cdots,a'_{2N})\prod^{2N}_{j=1}{\rm pf}(d_0,a_j)
\end{eqnarray}
where
\begin{eqnarray}\label{sdyo-57}
{\rm pf}(a'_j,a'_k)=\delta_{2N+1-j,k}\frac{1}{{\rm pf}(d_0,a_j){\rm pf}(d_0,a_{2N+1-j})}+\frac{p_j-p_k}{p_jp_k-1},\ \ j<k.
\end{eqnarray}
Thus, if $p_j$ satisfies the constraint condition:
\begin{eqnarray}\label{sdyo-58}
&&\nonumber\hspace{-2cm}\sum^M_{\mu=1}s^{(\mu)}\left( \frac{p_j-\alpha^{(\mu)}}{1-\alpha^{(\mu)}p_j}-\frac{1-\alpha^{(\mu)}p_j}{p_j-\alpha^{(\mu)}}
+ \frac{p_{2N+1-j}-\alpha^{(\mu)}}{1-\alpha^{(\mu)}p_{2N+1-j}}-\frac{1-\alpha^{(\mu)}p_{2N+1-j}}{p_{2N+1-j}-\alpha^{(\mu)}} \right)\\
&&=p_j-\frac{1}{p_j}+p_{2N+1-j}-\frac{1}{p_{2N+1-j}},
\end{eqnarray}
i.e.,
\begin{eqnarray}\label{sdyo-59}
&&\nonumber\hspace{-2.2cm}\sum^M_{\mu=1}s^{(\mu)}\left(\alpha^{(\mu)}{-}\frac{1}{\alpha^{(\mu)}}\right)\left[\frac{1}{(1{-}\alpha^{(\mu)}p_j)(1{-}\frac{1}{\alpha^{(\mu)}}p_{2N+1-j})}
+\frac{1}{(1{-}\alpha^{(\mu)}p_{2N+1-j})(1{-}\frac{1}{\alpha^{(\mu)}}p_{j})}\right]\\
&&=\frac{1}{p_j}+\frac{1}{p_{2N+1-j}},
\end{eqnarray}
then we have
\begin{eqnarray}\label{sdyo-60}
&& \partial_t\tau^{l^{(1)}\cdots l^{(M)}}_n=\sum^M_{\mu=1}s^{(\mu)}\partial_{y^{(\mu)}}\tau^{l^{(1)}\cdots l^{(M)}}_n.
\end{eqnarray}
Applying the above relation (\ref{sdyo-60}) to Eq. (\ref{sdyo-49}), we have
\begin{eqnarray}\label{sdyo-61}
\hspace{-1cm}[D_t-\sum^M_{\mu=1}s^{(\mu)} (\alpha^{(\mu)}-\frac{1}{\alpha^{(\mu)}})]f_{n+1}\cdot f_n=-\sum^M_{\mu=1}s^{(\mu)}(\alpha^{(\mu)}g^{(\mu)}_{n+1}h^{(\mu)}_n-\frac{1}{\alpha^{(\mu)}}g^{(\mu)}_{n}h^{(\mu)}_{n+1}).
\end{eqnarray}

Furthermore, imposing the complex conjugate conditions
\begin{eqnarray}\label{sdyo-62}
s^{(\mu)}=-{\rm i}\frac{c^{(\mu)}}{2},\ \ p_{2N+1-j}=\bar{p}_j,\ \ \xi^{(\mu)}_{p_{2N+1-j,0}}=\bar{\xi}^{(\mu)}_{j,0}+{\rm i}\frac{\pi}{2},\ \
\mbox{for} N\geq j\geq 1,
\end{eqnarray}
and requiring $y^{(\mu)}$ being pure imaginary, $|\alpha^{(\mu)}|=1$, one can have the function $h^{(\mu)}_n=\bar{g}^{(\mu)}_n$.

Consequently, Eqs.({\ref{sdyo-47}}),(\ref{sdyo-48}) and (\ref{sdyo-61}) become Eqs.({\ref{sdyo-52}})--(\ref{sdyo-54}).
In summary, the following Theorem holds:

\textbf{Theorem 4.1} The bilinear equations ({\ref{sdyo-52}})--(\ref{sdyo-54}) with $|\alpha^{(\mu)}|=1$ are satisfied by
\begin{eqnarray}\label{sdyo-63}
&& f_n=\tau^{0\cdots0}_n,\ \
 g_n^{(\mu )}  = \tau _n^{0...\mathop {\mathop {1} \limits^{\smile}}\limits^{\mu} ...0},\ \
 h_n^{(\mu )}  = \tau _n^{0...\mathop {\mathop {-1} \limits^{\smile}}\limits^{\mu} ...0},
\end{eqnarray}
where the tau function $\tau^{l^{(1)}\cdots l^{(M)}}_n$ is the Pfaffian $\tau^{l^{(1)}\cdots l^{(M)}}_n={\rm pf}(a_1,a_2,\cdots,a_{2N})$,
whose elements are defined by
\begin{eqnarray*}
&& {\rm pf}(a_j,a_k)=\delta_{2N+1-j,k}+\frac{p_j-p_k}{p_jp_k-1}{\rm pf}(d_0,a_j){\rm pf}(d_0,a_k),\\
&& {\rm pf}(d_l,a_j)=p^{n+l}_j\prod^M_{\mu=1}\left( \frac{p_j-\alpha^{(\mu)}}{1-\alpha^{(\mu)}p_j} \right)^{l^{(\mu)}}\exp(\xi_j),\\
&& \xi_j=(p_j-\frac{1}{p_j})t +\xi_{j,0}.
\end{eqnarray*}
Here $p_j$,  $\xi_{j0}$ are constants satisfying the following constraints.
\begin{eqnarray}\label{sdyo-64}
 &&-\sum^M_{\mu=1}{\rm i}\frac{c^{(\mu)}}{2}\left(\alpha^{(\mu)}-\bar{\alpha}^{(\mu)}\right)\Xi(\alpha^{(\mu)},p_j,p_{2N+1-j})=\frac{1}{p_j}+\frac{1}{p_{2N+1-j}},
\end{eqnarray}
with
\begin{eqnarray*}
&&\hspace{-2.2cm} \Xi(\alpha^{(\mu)},p_j,p_{2N{+}1{-}j})=\frac{1}{(1{-}\alpha^{(\mu)}p_j)(1{-}\bar{\alpha}^{(\mu)}p_{2N{+}1{-}j})} {+}\frac{1}{(1{-}\alpha^{(\mu)}p_{2N{+}1{-}j})(1{-}\bar{\alpha}^{(\mu)}p_{j})},
\end{eqnarray*}
and $p_{2N+1-j}=\bar{p}_j$, $\xi^{(\mu)}_{2N+1-j,0}=\bar{\xi}^{(\mu)}_{j,0}+{\rm i}\frac{\pi}{2}$ for $N\geq j\geq 1$.

In the following, we will illustrate one  and two dark soliton solutions for $M=2$.

\textbf{One-soliton solution:} By taking $N=1$ in (\ref{sdyo-63}),
we get the tau functions for the one-soliton solution:
\begin{eqnarray}
\label{sdyo-65}&& f_n=1+{\rm i}\frac{p_1-\bar{p_1}}{p_1\bar{p_1}-1}(p_1\bar{p_1})^n\exp(\xi_1+\bar{\xi}_1),\\
\label{sdyo-66}&& g^{(1)}_n=1+{\rm i}\frac{p_1-\bar{p_1}}{p_1\bar{p_1}-1}\frac{(p_1-\alpha^{(1)})(\bar{p_1}-\alpha^{(1)})}{(1-\alpha^{(1)} p_1)(1-\alpha^{(1)}\bar{p_1})}(p_1\bar{p_1})^n\exp(\xi_1+\bar{\xi}_1),\\
\label{sdyo-67}&& g^{(2)}_n=1+{\rm i}\frac{p_1-\bar{p_1}}{p_1\bar{p_1}-1}\frac{(p_1-\alpha^{(2)})(\bar{p_1}-\alpha^{(2)})}{(1-\alpha^{(2)} p_1)(1-\alpha^{(2)}\bar{p_1})}(p_1\bar{p_1})^n\exp(\xi_1+\bar{\xi}_1),
\end{eqnarray}
where $\xi_1=(p_1-\frac{1}{p_1})t+\xi_{1,0}$
and $p_1$ is a complex constant satisfying
\begin{eqnarray}\label{sdyo-68}
 &&\hspace{-2.3cm}{-}\sum^2_{\mu=1}{\rm i}\frac{c^{(\mu)}}{2}\left(\alpha^{(\mu)}{-}\bar{\alpha}^{(\mu)}\right)[\frac{1}{(1{-}\alpha^{(\mu)}p_1)(1{-}\bar{\alpha}^{(\mu)}\bar{p}_1)} {+} \frac{1}{(1{-}\alpha^{(\mu)}\bar{p}_1)(1{-}\bar{\alpha}^{(\mu)}p_1)}]=\frac{1}{p_1}{+}\frac{1}{\bar{p}_1}.
\end{eqnarray}

In order to avoid the singularity, the condition ${\rm i}\frac{p_1-\bar{p_1}}{p_1\bar{p_1}-1}>0$ need to be satisfied. Further, the above tau functions lead to the one-soliton solution as follows
\begin{eqnarray}
&& S^{(1)}_n=\frac{1}{2}e^{ {\rm i}(n\varphi_1+\frac{n}{2}\pi+2t\sin\varphi_1)} \left[1+e^{2{\rm i}\phi_1}-(1-e^{2{\rm i}\phi_1})\tanh(\xi'_{1R}+\theta_0) \right],\\
&& S^{(2)}_n=\frac{1}{2}e^{{\rm i}(n\varphi_2+\frac{n}{2}\pi+2t\sin\varphi_2)} \left[1+e^{2{\rm i}\phi_2}-(1-e^{2{\rm i}\phi_2})\tanh(\xi'_{1R}+\theta_0) \right],\\
&& L_n= \frac{(p_1\bar{p}_1-1)^2}{2p_1\bar{p}_1} {\rm sech}^2(\xi'_{1R}+\theta_0),
\end{eqnarray}
where $\xi'_1=\xi'_{1R}+{\rm i}\xi'_{1I}=n\ln({\rm i}p_1)+\xi_1$, $\exp({\rm i}\varphi_1)=\alpha^{(1)}$, $\exp({\rm i}\varphi_2)=\alpha^{(2)}$, $\exp(2\theta_0)={\rm i}\frac{p_1-\bar{p_1}}{p_1\bar{p_1}-1}$,
$\exp(2{\rm i}\phi_1)=\frac{(p_1-\alpha^{(1)})(\bar{p_1}-\alpha^{(1)})}{(1-\alpha^{(1)} p_1)(1-\alpha^{(1)}\bar{p_1})}$,
$\exp(2{\rm i}\phi_2)=\frac{(p_1-\alpha^{(2)})(\bar{p_1}-\alpha^{(2)})}{(1-\alpha^{(2)} p_1)(1-\alpha^{(2)}\bar{p_1})}$ and the parameter $p_1$ is determined by Eq. (\ref{sdyo-68}).
For the dark soliton in the SW components $S^{(1)}_n$ and $S^{(2)}_n$, their intensities approach 1 as $n\rightarrow\pm\infty$, and
the intensities of the center of the solitons read $\cos\phi_1$ and $\cos\phi_2$.
The real quantity $\frac{(p_1\bar{p}_1-1)^2}{2p_1\bar{p}_1}$ denotes the amplitude of the soliton in the LW component.
We illustrate single dark soliton for the choice of the nonlinearity coefficients $(c^{(1)},c^{(2)})=(1,-2)$ in Figure 3.
The parameters are chosen as $\alpha^{(1)}=\frac{\sqrt{2}}{2}+\frac{\sqrt{2}}{2}{\rm i}$, $\alpha^{(2)}=\frac{1}{2}+\frac{\sqrt{3}}{2}{\rm i}$, $p_1=1-0.7796{\rm i}$ and $\xi_{1,0}=0$.

\begin{figure}[!htbp]
\centering
{\includegraphics[height=1.8in,width=2.4in]{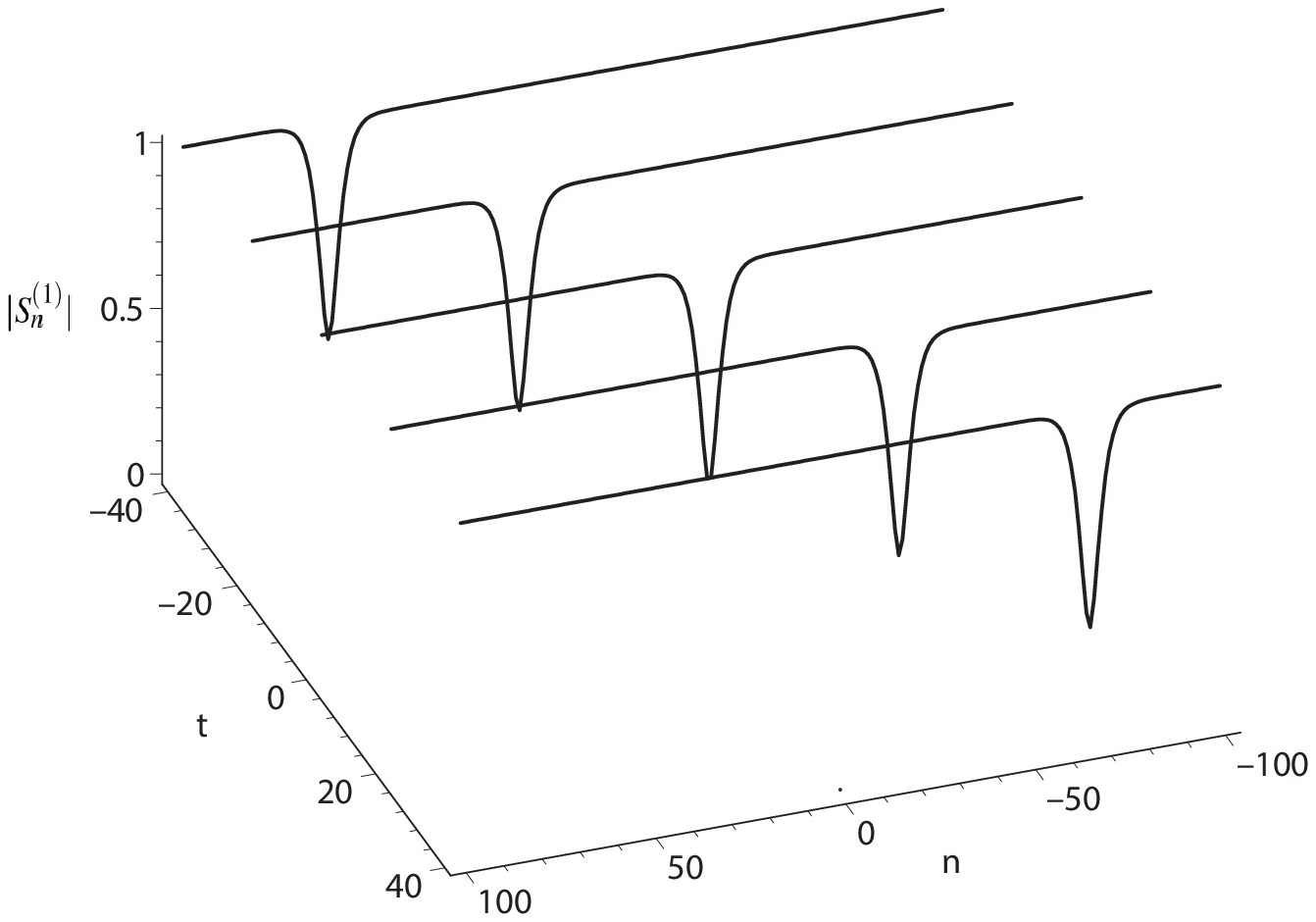}}
{\includegraphics[height=1.8in,width=2.4in]{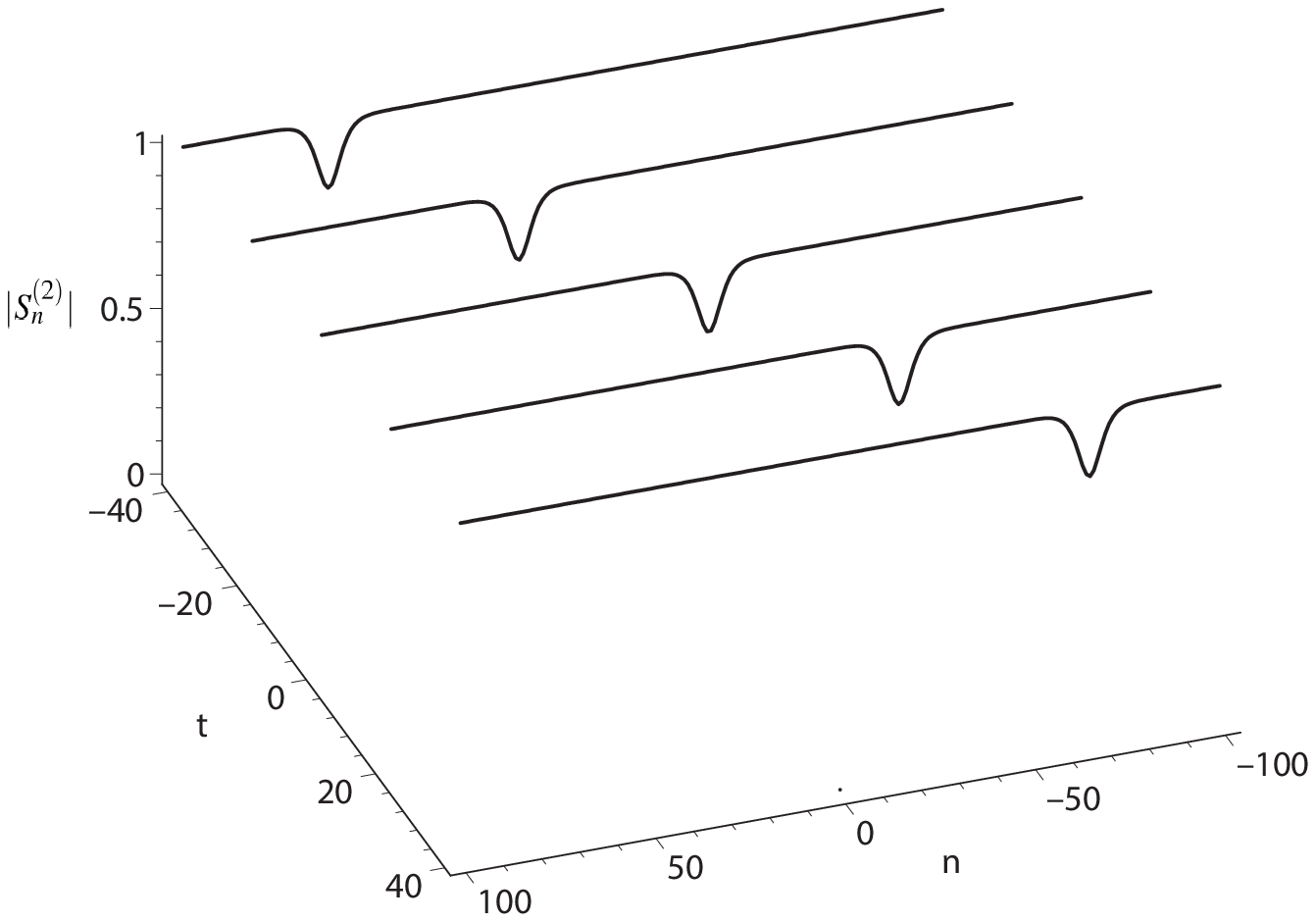}}
{\includegraphics[height=1.8in,width=2.4in]{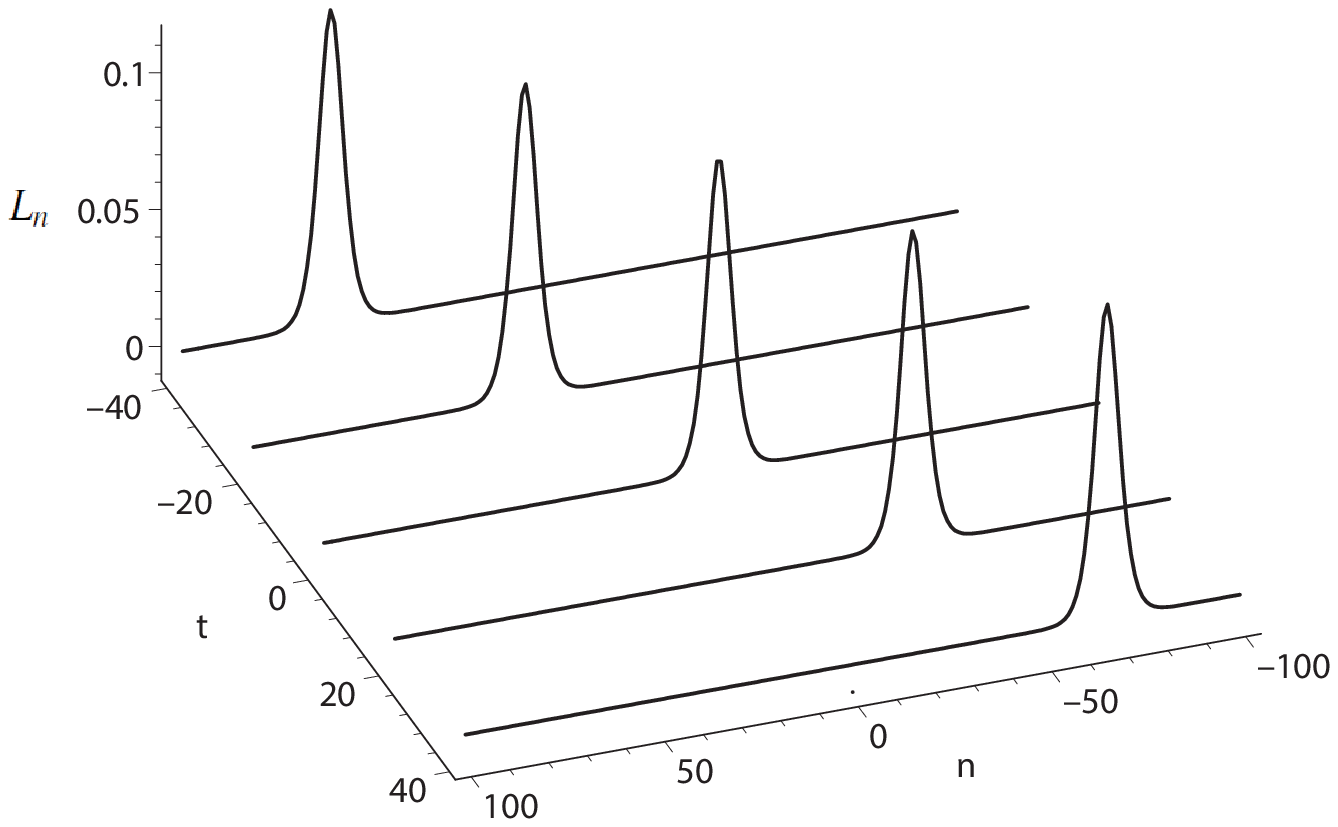}}
\caption{ The profiles of evolutions of one-soliton solution (dark soliton for SW components).
%with the parameters $
%(c^{(1)},c^{(2)})=(1,-2)$, $\xi'_{1,0}=0$, $\alpha^{(1)}=\frac{\sqrt{2}}{2}+\frac{\sqrt{2}}{2}{\rm i}$, $\alpha^{(2)}=\frac{1}{2}+\frac{\sqrt{3}}{2}{\rm i}$ and $p_1=1-0.7796{\rm i}$.
}
\end{figure}

\textbf{Two-soliton solution:} By taking $N=2$ in (\ref{sdyo-63}),
we get the tau functions for the two-soliton solution:
\begin{eqnarray}
\label{sdyo-69}&& f_n=1+E_1+E_2+\left|\frac{(p_1-p_2)(p_1-\bar{p}_2)}{(p_1p_2-1)(p_1\bar{p}_2-1)} \right|^2E_1E_2,\\
\label{sdyo-70}&& g^{(1)}_n=1+A_1E_1+A_2E_2+\left|\frac{(p_1-p_2)(p_1-\bar{p}_2)}{(p_1p_2-1)(p_1\bar{p}_2-1)} \right|^2A_1A_2E_1E_2,\\
\label{sdyo-71}&& g^{(2)}_n=1+B_1E_1+B_2E_2+\left|\frac{(p_1-p_2)(p_1-\bar{p}_2)}{(p_1p_2-1)(p_1\bar{p}_2-1)} \right|^2B_1B_2E_1E_2,,
\end{eqnarray}
where
\begin{eqnarray*}
&& E_j={\rm i}\frac{p_j-\bar{p}_j}{p_j\bar{p}_j-1}(p_j\bar{p}_j)\exp(\xi_i+\bar{\xi}_j),\\
&& A_j=\frac{(p_j-\alpha^{(1)})(\bar{p_j}-\alpha^{(1)})}{(1-\alpha^{(1)} p_j)(1-\alpha^{(1)}\bar{p_j})},\ \
B_j=\frac{(p_j-\alpha^{(2)})(\bar{p_j}-\alpha^{(2)})}{(1-\alpha^{(2)} p_j)(1-\alpha^{(2)}\bar{p_j})},
\end{eqnarray*}
and $\xi_j=(p_j-\frac{1}{p_j})t+\xi_{j,0}$
and $p_j$ are complex constants satisfying
\begin{eqnarray}\label{sdyo-72}
 &&\hspace{-2.3cm}{-}\sum^2_{\mu=1}{\rm i}\frac{c^{(\mu)}}{2}\left(\alpha^{(\mu)}{-}\bar{\alpha}^{(\mu)}\right)[\frac{1}{(1{-}\alpha^{(\mu)}p_j)(1{-}\bar{\alpha}^{(\mu)}\bar{p}_j)} {+} \frac{1}{(1{-}\alpha^{(\mu)}\bar{p}_j)(1{-}\bar{\alpha}^{(\mu)}p_j)}]=\frac{1}{p_j}{+}\frac{1}{\bar{p}_j},
\end{eqnarray}
for $j=1,2$.
Figure 4 exhibits such a two-soliton interaction with the parameters as $
(c^{(1)},c^{(2)})=(\frac{1}{9},-\frac{1}{8})$, $\alpha^{(1)}=\frac{\sqrt{2}}{2}+\frac{\sqrt{2}}{2}{\rm i}$, $\alpha^{(2)}=\frac{1}{2}-\frac{\sqrt{3}}{2}{\rm i}$, $p_1=1-0.6714{\rm i}$, $p_2=0.5-1.5356{\rm i}$ and $\xi'_{1,0}=\xi'_{2,0}=0$.

\begin{figure}[!htbp]
\centering
{\includegraphics[height=1.8in,width=2.4in]{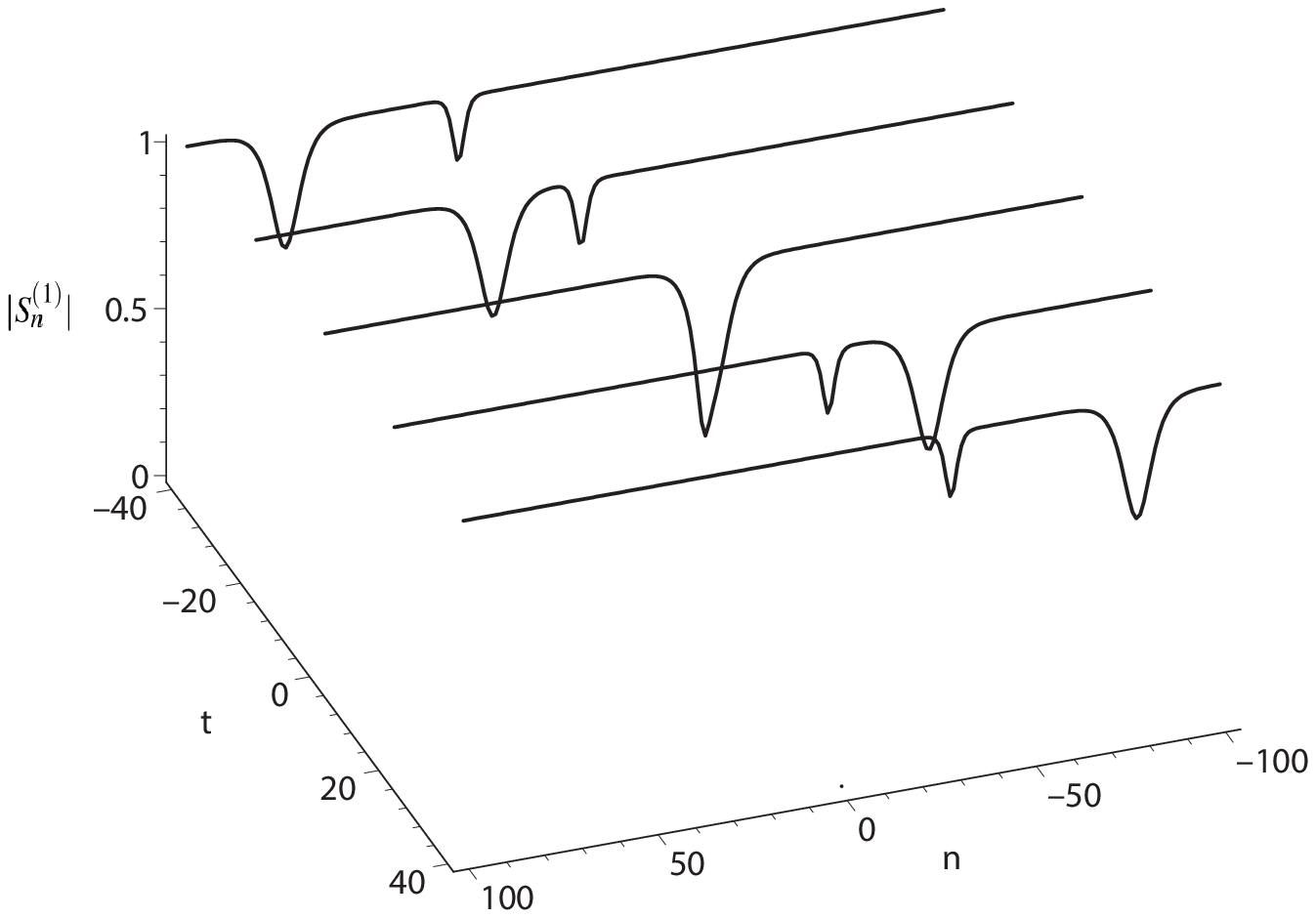}}
{\includegraphics[height=1.8in,width=2.4in]{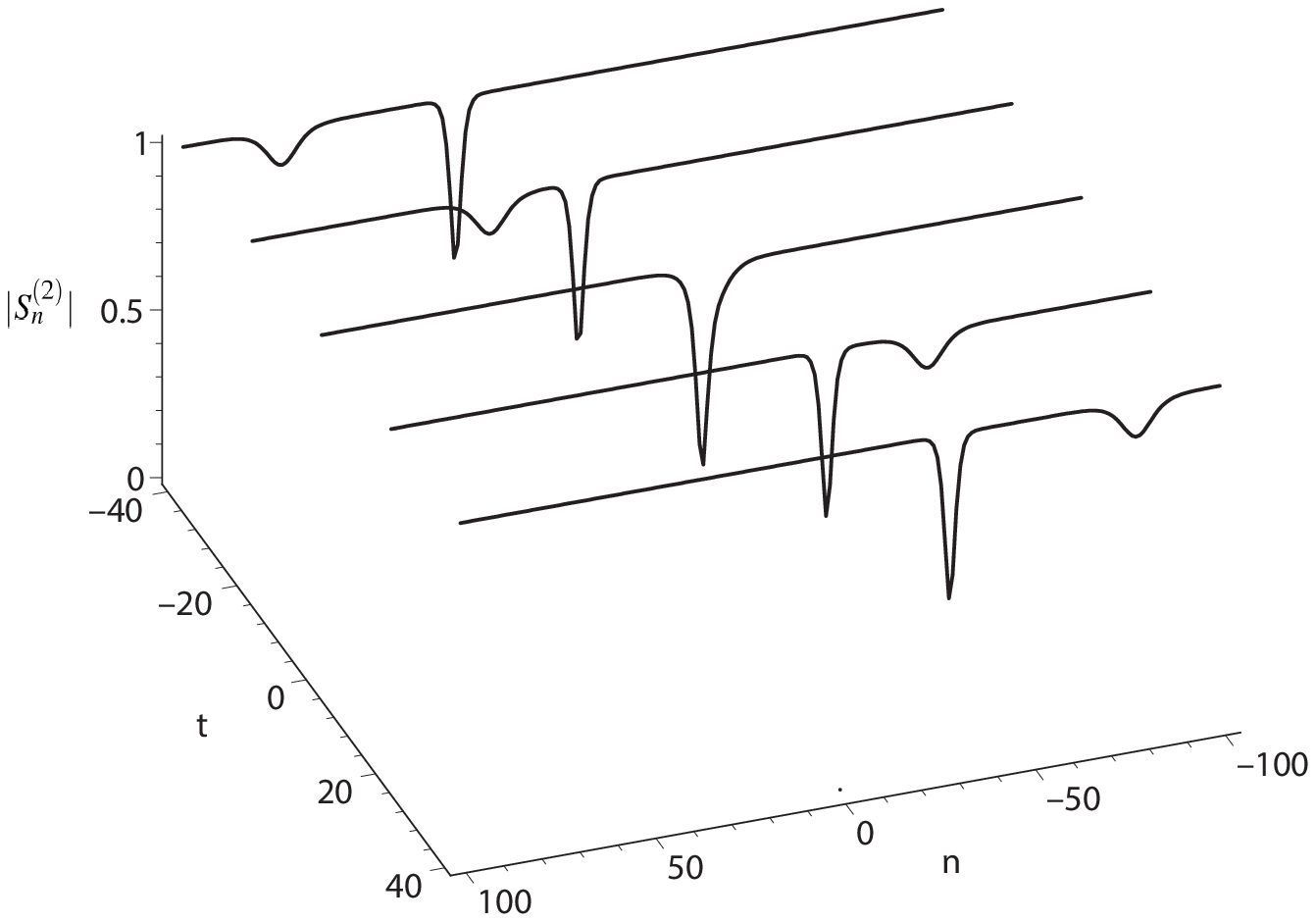}}
{\includegraphics[height=1.8in,width=2.4in]{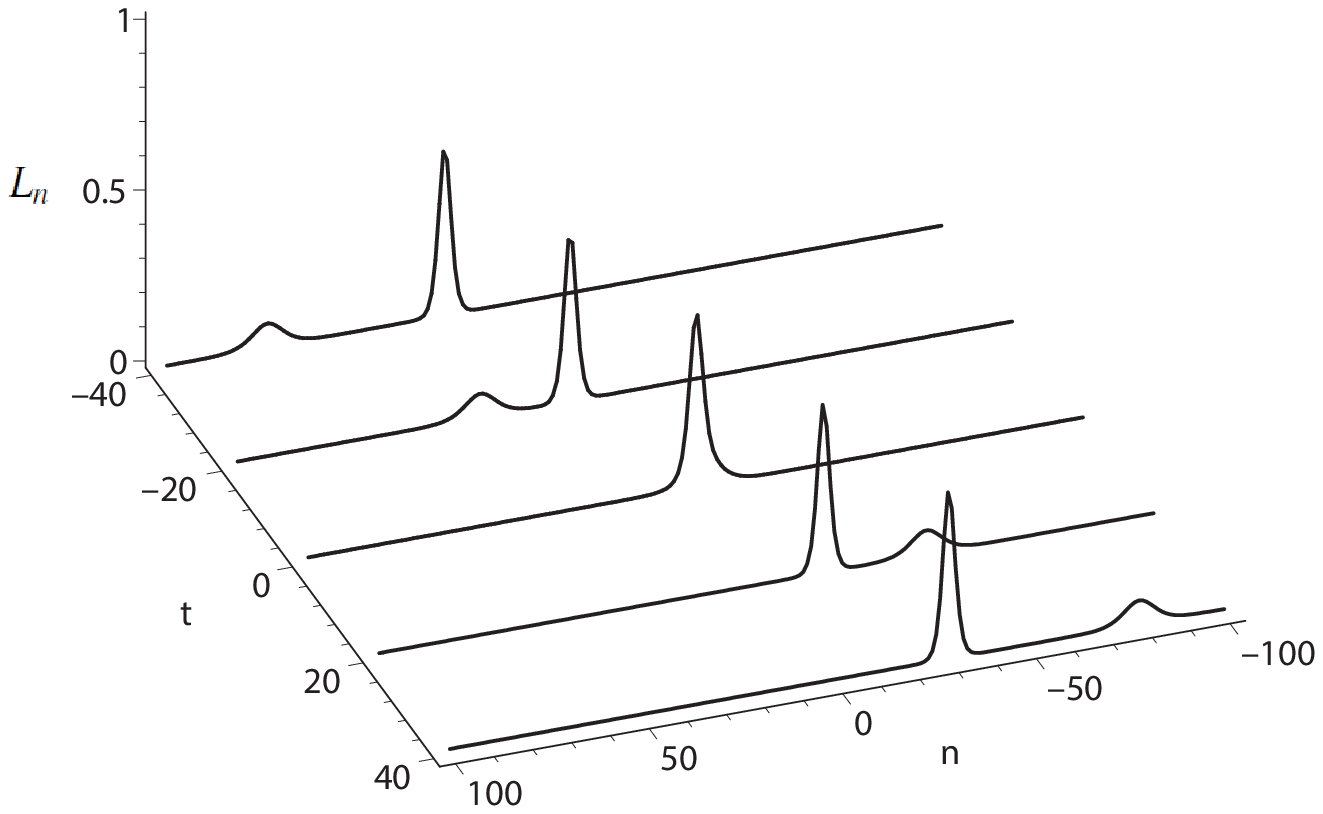}}
\caption{ The profiles of evolutions of two-soliton solution (dark soliton for SW components).
%with the parameters $
%(c^{(1)},c^{(2)})=(\frac{1}{9},-\frac{1}{8})$, $\xi'_{1,0}=\xi'_{2,0}=0$, $\alpha^{(1)}=\frac{\sqrt{2}}{2}+\frac{\sqrt{2}}{2}{\rm i}$, $\alpha^{(2)}=\frac{1}{2}-\frac{\sqrt{3}}{2}{\rm i}$ and $p_1=1-0.6714{\rm i}$, $p_2=0.5-1.5356{\rm i}$.
}
\end{figure}

\section{Conclusion and discussions}
%!
In the present paper, we construct an integrable semi-discrete analogue of the coupled YO system
by using Hirota's bilinear method. Moreover,
both the bright and dark soliton solutions in terms of pfaffians are derived
based on the B\"{a}cklund transformations of the semi-discrete BKP hierarchy.
As far as we are aware of, it is the first time to propose an integrable semi-discrete YO system.
%After finishing this paper, we have noted that the recently published paper \cite{yu2015dynamics} has constructed a differential-difference integrable (2+1)-dimensional system and investigated the bright soliton solution.
%Actually, this semi-discrete system is equivalent to a two-dimensional version of the single semi-discrete YO system (only including one SW component).
There are several topics left for the study of the semi-discrete YO system, which are listed below:\begin{enumerate}
       \item The Lax pair, the bi-Hamiltonian structure  and the conservation laws of this newly proposed semi-discrete YO system are not clear and deserve to be explored.
       \item How to construct the inverse scattering transform scheme for the semi-discrete YO system is also an interesting problem.
       \item It is of important to construct the mixed-type soliton solutions with both the bright and dark soliton for the vector semi-discrete YO system.
      \end{enumerate}

\section*{Acknowledgments}
J.C. appreciates the support by the China Scholarship
Council. The project is supported by the Global Change
Research Program of China (No. 2015CB953904), National
Natural Science Foundation of China (Grant Nos. 11275072,
11435005, and 11428102), Research Fund for the Doctoral
Program of Higher Education of China (No.
20120076110024), The Network Information Physics Calculation
of basic research innovation research group of China
(Grant No. 61321064), Shanghai Collaborative Innovation
Center of Trustworthy Software for Internet of Things (Grant
No. ZF1213), Shanghai Minhang District talents of high level
scientific research project and CREST, JST.

\section*{References}
%\bibliographystyle{unsrt}
%
%%\bibliography{ref}
%\bibliography{C:/Users/junchaochen/Desktop/ref}

\end{document}